\documentclass[preprint,12pt]{elsarticle}

\biboptions{sort&compress}

\usepackage{subcaption}
\usepackage{threeparttablex}

\usepackage{amsmath,
            amssymb,
            mathtools,
            amsthm,
            bm,
            mathrsfs,
            stmaryrd,
            upgreek,
            textcomp,
            cancel,
            stackengine,
            eqlist,
            graphicx,
            xfrac
}

\theoremstyle{definition}
\newtheorem{remark}{Remark}
\newtheorem{problem}{Problem}

\usepackage[np,autolanguage]{numprint}
\usepackage{
            multirow,
            float,
            rotating,
            afterpage,
            url%
}

\usepackage{nth}

\usepackage{ifthen}
\usepackage{calc}
\usepackage{xspace}

\usepackage{comment}
\excludecomment{tobedeleted}

\usepackage{tikz}
\newcommand*\circd[1]{%
  \begin{tikzpicture}
    \node[draw,circle,inner sep=0.5pt]{\scriptsize #1};
  \end{tikzpicture}}

\usepackage{acronym}
\acrodef{CRF}[CRF]{constitutive response function}
\acrodef{SI}[SI]{International System of Units}
\acrodef{BVP}[BVP]{boundary value problem}
\acrodef{FEM}[FEM]{finite element method}
\acrodef{EIT}[EIT]{extension, inflation, and torsion}
\acrodef{ALE}[ALE]{arbitrary {L}agrangian--{E}ulerian}

\acrodef{CNS}[CNS]{central nervous system}
\acrodef{CSF}[CSF]{cerebrospinal fluid}
\acrodef{IF}[IF]{interstitial fluid}
\acrodef{ECS}[ECS]{extracellular space}
\acrodef{SAS}[SAS]{subarachnoid space}
\acrodef{PVS}[PVS]{perivascular space}
\acrodef{BBB}[BBB]{blood-brain barrier}
\acrodef{WO}[WO]{Wang and Olbricht}
\acrodef{CMPH}[CMPH]{\protect\comsol}

\acrodef{BDF}[BDF]{backward differentiation formula}

\newcommand{\scdot}{\ensuremath{\!\cdot\!}}
\newcommand{\Grad}[2][nonsense]{%
\ifthenelse{\equal{#1}{nonsense}}%
{\nabla#2}%
{\nabla_{#1}\,#2}%
}
\newcommand{\Div}[2][nonsense]{%
\ifthenelse{\equal{#1}{nonsense}}%
{\nabla\scdot#2}%
{\nabla_{#1}\scdot#2}%
}
\newcommand{\Lin}[2][nonsense]{%
\ifthenelse{\equal{#1}{nonsense}}%
{\ensuremath{\mathscr{L}(#2)}}%
{\ensuremath{\mathscr{L}(#1,#2)}}%
}

\renewcommand{\d}[2][nonsense]{%
\ifthenelse{\equal{#1}{nonsense}}%
{\ensuremath{\mathrm{d}#2}}%
{\ensuremath{\mathrm{d}^{#1}#2}}
}
\newcommand{\bv}[1]{\boldsymbol{\mathbf{#1}}}

\newcommand{\ts}[1]{\text{$\mathsf{#1}$}}
\newcommand{\transpose}[1]{#1^{\scriptscriptstyle\mathrm{T}}}
\newcommand{\inversetranspose}[1]{\ensuremath{#1^{\scriptscriptstyle-\mathrm{T}}}}
\newcommand{\order}[3][nonsense]{%
\ifthenelse{\equal{#1}{nonsense}}%
{\ensuremath{#2^{\scriptscriptstyle(#3)}}}%
{\ensuremath{#2^{\scriptscriptstyle#1(#3)}}}%
}

\newcommand{\s}{\ensuremath{\mathrm{s}}}
\newcommand{\f}{\ensuremath{\mathrm{f}}}
\newcommand{\flt}{\ensuremath{\mathrm{flt}}}

\newcommand{\g}{\ensuremath{\mathrm{g}}}

\newcommand{\comsol}{COMSOL Multiphysics\textsuperscript{\textregistered}\xspace}

\makeatletter
\def\ps@pprintTitle{%
  \let\@oddhead\@empty
  \let\@evenhead\@empty
  \def\@oddfoot{\reset@font\hfil\thepage\hfil}
  \let\@evenfoot\@oddfoot
}
\makeatother

\begin{document}

\begin{frontmatter}
\title{Refinement of a Poroelastic Model for Zero Porosity: Finite Element Implementation and Investigation of Fluid Mechanics in the Perivascular Space}

\author[1,2]{Mohammad Jannesari}\ead{mbj5423@psu.edu}
\author[1,2,3]{Beatrice Ghitti}\ead{beatrice.ghitti@auckland.ac.nz}
\author[1,2]{Bruce J. Gluckman}\ead{bjg18@psu.edu}
\author[1,2]{Francesco Costanzo\corref{cor1}}\ead{fxc8@psu.edu}

\affiliation[1]{organization={Center for Neural Engineering},
city={University Park}, postcode={PA 16802}, country={USA}}
\affiliation[2]{organization={Engineering Science and Mechanics Department},
addressline={212 EES Bldg.},
city={University Park}, postcode={PA 16802}, country={USA}}
\affiliation[3]{organization={Auckland Bioengineering Institute},
addressline={The University of Auckland}, city={Auckland}, postcode={1010}, country={New Zealand}}

\cortext[cor1]{Corresponding author.}

\begin{abstract}
In conventional formulations of poroelasticity, when the porosity approaches zero or vanishes in some parts of the poroelastic domain, if only temporarily, the governing equations degenerate to those for the solid phase thereby inhibiting a suitable determination of the fluid velocity field. To address this challenge, we reformulated a poroelastic model based on mixture theory to accommodate scenarios with zero porosity. We verified our model using the method of manufactured solutions and demonstrated its ability to handle extreme conditions in a sample test problem. As an application of our framework, we investigated peristaltic flow in the perivascular space of a penetrating arteriole in brain. Our analysis revealed that some literature-suggested parameters can drive the model to predict extreme non-physiological conditions. We further demonstrated that these extreme conditions can be somewhat mitigated by accounting for the deformation of the surrounding brain tissue.
\end{abstract}


\end{frontmatter}

\section{Introduction}

The brain is housed within the skull and immersed in \ac{CSF}, which circulates around the brain and spinal cord \cite{Kucewicz2007Functional-tiss,TarasoffConway2015Clearance-syste,Abbott2018The-Role-of-Brain-0,Hladky2018Elimination-of-,Hladky2022The-glymphatic-}. The brain deformation is hypothesized to assist in this circulation by promoting the removal of metabolic waste products from the \ac{CNS} \cite{Sloots2020CardiacAndRespi}. In addition to the large-scale circulation of \ac{CSF}, evidence suggests that \ac{IF} also flows through the brain tissue itself and exchanges metabolites and other solutes with the broader \ac{CSF} system (cf.\ \cite{Iliff2012A-Paravascular-0, Iliff2013Brain-wide-Path0, Iliff2013Is-There-a-Cere-0, Iliff2013Cerebral-Arteri-0, Nedergaard2013Garbage-Truck-o, Rasmussen2022Fluid-transport}). Understanding these fluid flows is critical to gaining insights into both normal brain physiology and the development of pathological conditions such as neurodegenerative diseases \cite{Abbott2018The-Role-of-Brain-0, Nedergaard2013Garbage-Truck-o, Hladky2022The-glymphatic-}.
However, directly observing and quantifying these flows \emph{in vivo} is extraordinarily challenging via currently available experimental techniques. 
For example, the pressure differences driving \ac{CSF} through \ac{PVS} can be as small as $\np[mmHg]{0.01}$---$\np{100}$ times smaller than the approximately $\np[mmHg]{1}$ resolution of current instruments \cite{Oshio2004ReducedCerebros,Uldall2017AcetazolamideLo}---and are further disrupted by invasive skull access \cite{Mestre2018Aquaporin4Depen}.
As a result, physics-based mathematical modeling has become an attractive tool to evaluate competing hypotheses about the drivers and pathways of brain fluid flow \cite{Hladky2022The-glymphatic-}.

Our group has been actively engaged in this field, especially in exploring the dynamics of fluid flow and exchange between the brain parenchyma and the \ac{SAS}. In particular, \ac{CSF} flow and exchange in the \ac{PVS} have been investigated using computational modeling techniques, showing that, under idealized conditions, arterial pulsations alone may not suffice to drive flow within the arterial \ac{PVS} \cite{Kedarasetti2020Arterial-Pulsat, Kedarasetti2020Functional-Hype}.
A crucial insight from our work is the importance of accounting for the deformability of brain tissue when modeling these processes. In fact, we have shown that mechanical deformations affect the effective permeability of the brain parenchyma and influence how fluids can move through it \cite{Kedarasetti2022ArterialVasodil, Garborg2025Gut-,Costanzo2025Por}. 

In the course of our modeling effort, we encountered scenarios in which numerical methods struggled to adequately represent the physics. Specifically, under the action of periodic compressive loads, numerical experiments showed that deformation can strongly affect porosity so as to drive fluid out of some regions of the simulation's domain, thereby leaving these regions nearly or entirely depleted of fluid. These results revealed a limitation in our poroelastic models when dealing with extreme cases of near-zero porosity, pointing out the need for a more refined model that can handle and accurately represent an extreme condition like vanishing porosity. Pursuing the creation of such a model is not motivated merely by numerical considerations. We believe that a stronger rationale stems from the role of computational modeling in testing physiological hypotheses through the lens of physics. Since some of these hypotheses may initially be incomplete or oversimplified, it is valuable to identify when and why a model prediction departs significantly from physiologically relevant conditions. This allows us to correct or discard any nonphysical assumptions.

Motivated by these considerations, in this work we reformulate our poroelastic framework, previously adopted in several publications \cite{Kedarasetti2020Arterial-Pulsat, Kedarasetti2020Functional-Hype,Kedarasetti2022ArterialVasodil, Garborg2025Gut-}, to handle situations where deformation completely expels the fluid out from parts of a poroelastic body, resulting in regions of vanishing porosity. We implement the model numerically, verify it using the method of manufactured solutions, and identify stable finite element spaces suitable for these cases.
Although this study does not focus primarily on modeling brain physiology, here we revisit a flow problem simulating fluid flow in the \ac{PVS} surrounding a cerebral vessel, caused by peristaltic motion of the arterial walls, to demonstrate the relevance of our approach. This problem illustrates how assumptions and parameter values in a model can yield predictions that diverge from physiological expectations. In addition, our findings further reinforce our belief that accounting for tissue deformability and the physics of porosity dynamics are essential to reach consistent and meaningful conclusions.

In Section \ref{sec: traditional formulation}, we revisit the traditional formulation of poroelasticity based on mixture theory. The modifications to capture the vanishing-porosity behavior are introduced in Section \ref{sec: reformulation}. The weak form of the refined model is presented in Section \ref{sec: weak form}, where a verification test via the method of manufactured solutions and a local compression benchmark problem are also performed (Sections \ref{subsec:mms} and \ref{subsec:benchmark}, respectively). Finally, in Sections \ref{sec: PVS} we apply our model to the analysis of fluid flow in the \ac{PVS} by peristalsis.

\section{Traditional formulation: definitions and governing equations}\label{sec: traditional formulation}

Here we report the governing equations of a poroelastic body starting from mixture theory \cite{Bowen1976TheoryOfMixture}. 

\subsection{Constituent phases and mixture}
\begin{figure}[htb]
	\centering
	\includegraphics[width=0.6\linewidth]{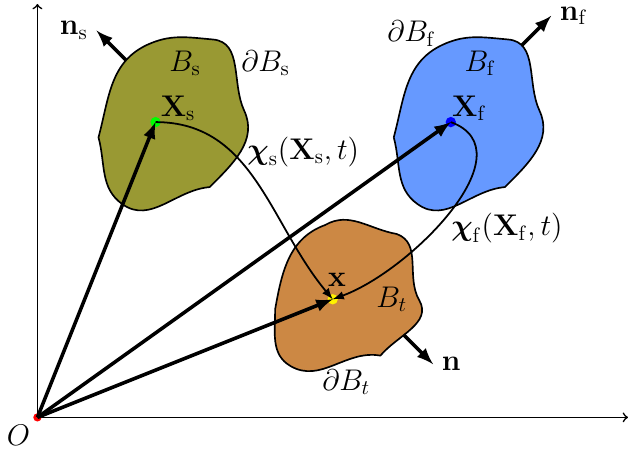}
	\caption{Current configuration $B_t$ of a biphasic mixture, along with the reference configurations of solid and fluid components, $B_\s$ and $B_\f$, respectively. Positions within $B_\s$ and $B_\f$ will be labeled as $\bv{X}_\s$ and $\bv{X}_\f$, respectively, while $\bv{x}$ signifies the position within $B_t$. The boundaries of $B_\s$, $B_\f$, and $B_t$ are $\partial B_\s$, $\partial B_\f$, and $\partial B_t$, respectively. These boundaries are oriented by their respective outward unit normal fields $\bv{n}_\s$, $\bv{n}_\f$, and $\bv{n}$.}
	\label{fig:configurations}
\end{figure}

We consider a saturated binary mixture of a solid and a fluid phase. At the current time $t$, the mixture occupies the subset $B_{t}$ (Fig.~\ref{fig:configurations})
of an all-containing Euclidean point space $\mathcal{E}^{d}$, with $d=1,2$ or $3$. The constituents of the mixture can coexist at each spatial point $\bv{x}\in B_{t}$. Subscripts $\s$ and $\f$ will be used to denote quantities pertaining to the solid and fluid phases, respectively.  As shown in Fig.~\ref{fig:configurations}, we posit that there exists a diffeomorphism 
$\bv{\chi}_a:B_a\times[0,T]\rightarrow B_t\subset \mathcal{E}^d$
which maps the reference configuration $B_a$ of component $a$ ($a=\s, \f$) to the mixture's current configuration $B_t$. The position of points in $B_{a}$ will be denoted by $\bv{X}_{a}$. For simplicity, the initial configurations of the components are assumed to coincide with their respective reference configurations and with $B_{t}|_{t = 0}$, and therefore with the initial configuration of the porous structure as a whole. We regard $\mathcal{E}^{d}$ as covered by a single Cartesian coordinate system. We denote by $\mathscr{V}$ the translation space of $\mathcal{E}^{d}$ and by $\mathscr{E}=\{\bv{e}_{1},\ldots,\bv{e}_{d}\}$ the orthonormal basis of $\mathscr{V}$ that is induced by the coordinate cover.

\subsection{Concerning the domains of functions}
The functions describing the physics of the system can have domains $B_{\s}$, $B_{\f}$, and $B_{t}$. Later, we will reformulate the equations to solve them numerically via the \ac{FEM} while adopting an \ac{ALE} scheme (cf.\ \cite{Fernandez2009The-Derivation-0}). Therefore, we will need to refer to the same physical quantity but expressed by different functions when said quantity is remapped over a specific domain. With this aim, and to avoid the proliferation of symbols, we adopt the short-hand notation in \cite{dellIsola2009Boundary-Condit-0}  to indicate the domain to which a function is being remapped.

The symbols $\circd{e}$, $\circd{s}$, and $\circd{f}$ indicate $B_t$, $B_{\s}$, and $B_{\f}$, respectively. With $a,b = \s,\f, t$, let $\varphi: B_{a}\to\mathscr{U}$ and $\zeta: B_{b}\to\mathscr{W}$ be smooth functions taking values in the vector spaces $\mathscr{U}$ and $\mathscr{W}$, respectively. Let $\bv{\chi}_{ab}: B_{a}\to B_{b}$ be a diffeomorphism between $B_{a}$ and $B_{b}$. Then, we denote by $\varphi^{\circd{b}}$ and $\zeta^{\circd{a}}$ the following functions
\begin{equation}
\label{eq: circd notation}
\varphi^{\circd{b}} \coloneqq \varphi \circ \bv{\chi}_{ab}^{-1}
\quad\text{and}\quad
\zeta^{\circd{a}}  \coloneqq \zeta \circ \bv{\chi}_{ab}.
\end{equation}

\subsection{Kinematics}
Following \cite{Bowen1976TheoryOfMixture,Costanzo2016Finite-Element-0}, for each constituent $a=\s, \f$, the displacement field, the deformation gradient, and its determinant are defined as, respectively,
\begin{align}
    \label{eq: disp definition}
		\bv{u}_{a}\left(\bv{X}_{a}, t\right) &\coloneqq\bv{\chi}_{a}\left(\bv{X}_{a}, t\right)-\bv{X}_{a},
        \\
        \label{eq: Fa definition}
		\ts{~F}_{a}\left(\bv{X}_{a}, t\right) &\coloneqq\frac{\partial \bv{\chi}_{a}\left(\bv{X}_{a}, t\right)}{\partial \bv{X}_{a}},
        \\
        \label{eq: Ja definition}
        J_{a}\left(\bv{X}_{a}, t\right)&\coloneqq \det \ts{F}_{a}\left(\bv{X}_{a}, t\right),
\end{align}
where we note that $\bv{u}_{a} = \bv{x} - \bv{X}_{a}$ since $\bv{x}=\bv{\chi}_{a}\left(\bv{X}_{a}, t\right)$.

The material velocity for the constituent $a$ is
\begin{equation}
	\bv{v}_{a}(\bv{x}, t)
    \coloneqq\left.\frac{\partial \bv{\chi}_{a}\left(\bv{X}_{a}, t\right)}{\partial t}\right|_{\bv{X}_{a}=\bv{\chi}_{a}^{-1}(\bv{x}, t)}
    =\left.\frac{\partial \bv{u}_{a}\left(\bv{X}_{a}, t\right)}{\partial t}\right|_{\bv{X}_{a}=\bv{\chi}_{a}^{-1}(\bv{x}, t)}. 
\end{equation}
We note that the domain of $\bv{v}_{a}$ is $B_{t}$. The material acceleration of phase $a$ is also a function with domain $B_{t}$ and is given by
\begin{equation}
\label{equ: material acceleration}
\bv{a}_{a} = \partial_{t} \bv{v}_{a} +
\Grad{\bv{v}_{a}}[\bv{v}_{a}],
\end{equation}
where the notation $\Grad \square[\diamond]$ denotes the action of the gradient of the quantity `$\square$'  onto the field `$\diamond$,' provided such an action is meaningful.  For future use we will denote the divergence operator by `$\Div{}$'.
\begin{remark}[Concerning $\Grad{}$ and $\Div{}$]
    The  gradient and divergence operators presume the use of an underlying metric, the latter defined over a specific domain. As functions can have different domains, the meaning and practical computation of the action of these operators cannot be divorced from the domain of said function. Therefore, as an example, letting $p:B_{t}\to\mathbb{R}$, the functions $\Grad{p}$ and $\Grad{(p^{\circd{s}})}$ are different from one another because the former is the gradient of $p$ over $B_{t}$ whereas the latter is the gradient over $B_{\s}$ of the function $p^{\circd{s}}$.
\end{remark}

The volume fraction of species $a$ will be denoted by $\phi_{a}(\bv{x},t)$. We assume that the porous medium is saturated, so we have $\phi_\s+\phi_\f=1$.

For fluid flow models in which the fluid is assumed incompressible (the fluid's mass density is a constant), the fluid's velocity flux through a surface can be used as a direct indicator of the corresponding mass flux. This fact justifies the introduction of flow indicators like the \emph{flow rate} in engineering fluid mechanics (cf.\ \cite{Cengel2017Fluid-Mechanics}). However, in the case of flow through a porous medium, the fluid's velocity flux can no longer be taken as a direct indicator of a corresponding mass flux through the porous medium. In our current modeling context, the information in question is linked to (\emph{i}) the velocity of the fluid \emph{relative} to that of the solid when (\emph{ii}) this velocity is appropriately scaled by the volume fraction of the fluid phase. Therefore, in porous flow models, especially when the phases are taken to be incompressible in their pure state, a new velocity field is introduced that can be used in a way similar to the true fluid velocity in pure fluid flow models. This velocity field is the \emph{filtration velocity}, which is defined as 
\begin{equation}
\label{eq: vflt def}
\bv{v}_\flt \coloneqq \phi_\f\left( \bv{v}_\f - \bv{v}_\s \right).
\end{equation}
In the Darcy flow problem, the filtration velocity is the primary unknown of the problem (cf., e.g., \cite{Whitaker1986Flow-in-Porous-,Wang2011Fluid-Mechanics-0}) and it vanishes when $\phi_{\f}\to\np{0}$. Furthermore, in the traditional form of the Darcy problem, it is the filtration velocity that is involved in the statement of boundary conditions (e.g., by setting the normal component of the filtration velocity to zero to characterize an impermeable boundary; see, e.g., \cite{Masud2007A-Stabilized-Mixed-0}).

\subsection{Governing equations}
We assume that the mixture's constituents are incompressible in their pure form. Enforcing this constraint along with the balance of mass in the absence of chemical reactions, we have
\begin{equation}
	\label{equ:mass_balance_component}
	\partial_{t} \phi_{a}+
    \Div{(\phi_{a} \bv{v}_{a})}= 0 \quad\Rightarrow\quad
\phi_{a} = (\phi_{R_{a}}/J_{a})^{\circd{e}},
\end{equation}
where $\phi_{R_{a}}:B_{a}\to\mathbb{R}$ is the referential volume fraction of component $a$ (cf. \cite{Costanzo2016Finite-Element-0}).
Summing the first of Eq.~\eqref{equ:mass_balance_component} 
for the solid and fluid components and considering the saturation constraint, we then have
\begin{equation}
	\label{equ:mass_balance_mixture}
        \Div
        {\left(
	\phi_{\s} \bv{v}_{\s} +
	\phi_{\f} \bv{v}_{\f}
	\right)}=0 \quad \text{in $B_t$}.
\end{equation}
We will refer to this equation as the \emph{continuity equation} as it plays a role practically identical to the continuity equation in the Stokes and Navier–Stokes problems.

For the solid phase, let $\ts{C}_{\s}$ denote the right Cauchy-Green strain tensor and $\bar{\ts{C}}_{\s}$ its corresponding isochoric component, i.e.,
\begin{equation}
\label{eq: Cs and Csbar defs}
\ts{C} = \transpose{\ts{F}}_{\s} \ts{F}_{\s}
\quad\text{and}\quad
\bar{\ts{C}}_\s=J_\s^{-2/3}\ts{C}_\s.
\end{equation}
We assume that the solid phase in its pure form is hyperelastic and we denote by $\hat{\Psi}_{\s}=\hat{\Psi}_{\s}\left(\bar{\ts{C}}_{\s}\right)$ its strain energy per unit volume. We then assume that the strain energy per unit volume of the solid's phase in its reference configuration is
\begin{equation}
	\left.\Psi_\s(\bv{X}_\s, t)\right.=\phi_{R_{\s}}\left(\bv{X}_{\s}\right) \hat{\Psi}_{\s}\left(\bar{\ts{C}}_{\s}\left(\bv{X}_{\s}, t\right)\right).
\end{equation}
In the current study, we select a neo-Hookean material model so that $\hat{\Psi}_{\s}=\tfrac{1}{2}\mu_\s\left( \bar{I}_1 -3 \right)$, where $\bar{I}_1$ represents the first principal invariant of $\bar{\ts{C}}_{\s}$ and $\mu_\s$ denotes the solid shear modulus.
The elastic and total Cauchy stress tensors for the solid phase are (cf.\ \cite{Bowen1980Incompressible-0})
\begin{equation}
	\label{equ:stress}
	\ts{T}^\mathrm{e}_\s=
    \biggl(
    2 \frac{\phi_{R_\s}}{J_\s} \ts{F}_{\s} \frac{\partial \hat{\Psi}_\s}{\partial \ts{C}_{\s}} \transpose{\ts{F}_{\s}}\biggr)^{\circd{e}}
	\quad\Rightarrow\quad
	\ts{T}_\s=-\phi_\s p \ts{I}+\ts{T}^\mathrm{e}_\s,
\end{equation}
where $p = p(\bv{x},t)$ is the pore pressure.

When surface tension effects are neglected in the porous medium, the interaction of the mixture's constituents is typically limited to a drag force proportional to the velocity of the fluid relative to the solid. Then, following \cite{Bowen1976TheoryOfMixture,Costanzo2016Finite-Element-0}, and enforcing the balance of momentum for the phases as well as the mixture, the equations of motion for the solid and fluid constituents are, respectively,
\begin{align}
\label{equ:eom_solid}
    \rho_{\s} (\bv{a}_{\s}-\bv{b}_{\s})+\phi_{\s} \Grad{p}
    -\phi_{\f}^{2} \frac{\mu_{\f}}{\kappa_{\s}}\left(\bv{v}_{\f}-\bv{v}_{\s}\right)-\Div{\ts{T}^\mathrm{e}_\s}
& =\bv{0} &&\text{in $B_{t}$},
\\
\label{equ:eom_fluid}
    	\rho_{\f}(\bv{a}_{\f}-\bv{b}_{\f})+\phi_{\f} \Grad{p}+\phi_{\f}^{2} \frac{\mu_{\f}}{\kappa_{\s}}\left(\bv{v}_{\f}-\bv{v}_{\s}\right)
    &=\bv{0} &&\text{in $B_{t}$},
\end{align}
where $\mu_\f$ is the dynamic viscosity of the fluid component, $\kappa_\s$ is the permeability of the solid component, $\bv{b}_a$ is the body force per unit mass of the constituent $a$, and $\rho_a$ is the mass density of constituent $a$ per unit volume of the mixture, given by
\begin{equation*}
\label{equ: mass density of a}
    \rho_a=\phi_a\rho^*_a,
\end{equation*}
where $\rho^*_a$ is the density of the constituent $a$ in its pure form.

Taken together, Eqs.~\eqref{equ:mass_balance_mixture}, \eqref{equ:eom_solid}, and~\eqref{equ:eom_fluid} constitute the governing equations of the system.

Often Eq.~\eqref{equ:eom_fluid} is presented in a slightly simplified form under the assumption that $\phi_{\f} > \np{0}$. In this case, one can divide Eq.~\eqref{equ:eom_fluid} by $\phi_{\f}$ and write
\begin{equation}
\label{equ:eom_fluid_simplified}
    \rho^*_{\f}(\bv{a}_{\f}-\bv{b}_{\f})+ \Grad{p}+\phi_{\f} \frac{\mu_{\f}}{\kappa_{\s}}\left(\bv{v}_{\f}-\bv{v}_{\s}\right)=\bv{0} \quad \text{in $B_{t}$},
\end{equation}
which is an equation often associated with Darcy flow, especially when it is assumed that the solid phase is stationary (cf.\ \cite{Wang2011Fluid-Mechanics-0}).

In other formulations, Eq.~\eqref{equ:eom_solid} is replaced by an  equation of motion for the mixture as a whole, obtained by summing Eqs.~\eqref{equ:eom_solid}  and~\eqref{equ:eom_fluid}:
\begin{equation}
\label{equ:eom_mixture}
	\rho_{\s}(\bv{a}_{\s}-\bv{b}_{\s})
	+\rho_{\f}(\bv{a}_{\f}-\bv{b}_{\f})+
	\Grad{p}-\Div{\ts{T}^\mathrm{e}_\s}
=\bv{0} \quad \text{in $B_{t}$}.
\end{equation}

\begin{remark}[The need for a modified model for vanishing porosity]
    Recalling the saturation constraint ($\phi_{\s}+\phi_{\f} = 1$) and assuming that every field describing the physics remains bounded, we observe that the limit as $\phi_{\f}\to \np{0}$ of Eq.~\eqref{equ:mass_balance_mixture} can be considered well defined. The same can be said for Eqs.~\eqref{equ:eom_solid} and~\eqref{equ:eom_mixture}. However, as $\phi_{\f}\to \np{0}$, both Eqs.~\eqref{equ:eom_fluid} and~\eqref{equ:eom_fluid_simplified} present some issues. Equation~\eqref{equ:eom_fluid} would become trivial altogether, whereas Eq.~\eqref{equ:eom_fluid_simplified} becomes undefined since its very derivation required a division by $\phi_{\f}$. Even if one were to take Eq.~\eqref{equ:eom_fluid_simplified} as \emph{the} equation of motion of the fluid (as opposed to having been derived from a parent statement), the limit for $\phi_{\f} \to \np{0}$ would present some difficulties as it would imply that the fluid's inertia does not vanish with the fluid's volume fraction and that the motion of the fluid is driven by the gradient of $p$, the latter having become a multiplier for the enforcement of the constraint needed to ensure that the motion of the solid is volume preserving.
\end{remark}

\begin{remark}[Concerning traction boundary conditions]
\label{remark: on tractions}
    The application of traction boundary conditions in mixture theory is a delicate subject (cf.\ \cite{Rajagopal1995Mechanics-of-Mixtures-0,Rajagopal1986On-Boundary-Conditions-for-a-Certain-0}). In this paper, we will not need a model to manage the partition of a traction distribution on the boundary of the mixture in corresponding traction distributions over individual phases. Hence, for later use, we simply denote by $\bv{s}$ the boundary traction on the mixture:
\begin{equation}
\label{equ: boundary traction def}
\bv{s} = \ts{T} \bv{n},
\end{equation}
where $\ts{T}$ is the total Cauchy stress acting on $\partial B_{t}$ and $\bv{n}$ is the outer unit normal orienting $\partial B_{t}$. For the model presented thus far, we have
\begin{equation}
\label{equ: total T}
\ts{T} = -p \, \ts{I} + \ts{T}_{\s}^{\mathrm{e}}.
\end{equation}
\end{remark}

\section{Reformulation for the case of vanishing porosity}\label{sec: reformulation}
In elasticity, the numerical analysis of incompressible or nearly incompressible materials---where the Poisson’s ratio approaches $\np{0.5}$---presents specific numerical challenges that must be addressed to obtain accurate solutions (see, e.g., \cite{Glowinski1982NumericalSoluti}).
Analogously, the common formulation of poroelasticity cannot handle the limit-case condition with $\phi_{\f}\to \np{0}$. 
If the porosity $\phi_\f$ approaches zero at some point $\bv{x}$ in the domain $B_{t}$, the equation of motion of the fluid, Eq.~\eqref{equ:eom_fluid}, becomes trivial at that point, leading to a loss of well-posedness and resulting in a singular problem.
It is well known that numerical failure in solving the governing equations does not require $\phi_{\f}=\np{0}$; the system may become unstable or ill-posed when $\phi_f$ is sufficiently small in certain regions of the domain. 
These observations justify the need to develop a refined version of the model, one that can deliver solutions, especially numerical solutions, even when $\phi_{\f}$ approaches zero over some subset of the solution's domain.

To formulate a modification of Eq.~\eqref{equ:eom_fluid}, we start with the observation that, in a continuum-level mixture theory, all phases coexist at all the spatial points occupied by the mixture. That is, there is no explicit recognition that the solid and fluid phases are otherwise separated so to occupy distinct regions within $B_{t}$. This said, we envision a microscopic picture in which this separation does exist. In this framework, we view the condition $\phi_{\f}\to \np{0}$ as describing the vanishing of the physical space within which the fluid particles can move. We also assume that, at microscopic scale, a no-slip condition exists at any surface in which the fluid comes into contact with the solid phase. Therefore, we posit that, as $\phi_\f \to \np{0}$, the fluid velocity $\bv{v}_\f$ approaches the solid velocity $\bv{v}_\s$, i.e., we posit that if $\phi_{\f}\to \np{0}$ at a point, $\|\bv{v}_{\f}-\bv{v}_{\s}\|\to \np{0}$ at that point. This assumption is inherently related to the question pertaining to the scaling of the permeability $\kappa_{\s}$ with the fluid porosity. This question has been addressed both empirically and theoretically, and the associated literature supports our assumption (cf.\ \cite{Yuan2023PorosityPermeab,OBrien2006TheEffectOfPore,Carman1939PermeabilityOfS,Gebart1992permeability}).

\begin{remark}[The filtration velocity and its dependence on porosity]
\label{remark: vflt and porosity}
Equation~\eqref{eq: vflt def} implies that $\|\bv{v}_{\flt}\|\to \np{0}$ when $\phi_{\f}\to \np{0}$. In light of the assumption we have made concerning $\|\bv{v}_{\f}-\bv{v}_{\s}\|$, we can say that, at those points where $\phi_{\f}\to \np{0}$,  $\|\bv{v}_{\flt}\| = o(\phi_{\f})$, that is, $\|\bv{v}_{\flt}\|$ goes to zero faster than $\phi_{\f}$.
\end{remark}

Given the above premise, in order to formulate a suitable modification of Eq.~\eqref{equ:eom_fluid},
we define an auxiliary variable $\eta=\eta(\phi_\f)$ as
\begin{equation}\label{equ:eta_def}
    \eta(\phi_\f) \coloneqq \frac{\mu_\f \phi_\f^2}{\kappa_\s(\phi_\f)},
\end{equation}
such that
\begin{equation}\label{equ:eta_cond}
    \lim_{\phi_\f\to 0} \frac{\phi_{\f}}{\eta(\phi_\f)} = 0,
\end{equation}
where $\mu_{\f}$ is still treated as a material constant.
The condition in Eq.~\eqref{equ:eta_cond} allows us to formalize a constitutive assumption concerning the permeability of the porous medium. Specifically, we assume that 
\begin{equation}
\label{equ: permeability assumption}
\kappa_\s(\phi_\f)=o({\phi_\f}).
\end{equation}
By dividing
Eq.~\eqref{equ:eom_fluid}
by $\eta(\phi_\f)$, then we have
\begin{equation}\label{equ:eom_fluid_mod}
    \frac{\phi_{\f} \rho^*_\f}{\eta(\phi_\f)} (\bv{a}_\f-\bv{b}_\f)+\frac{\phi_{\f}}{\eta(\phi_\f)} \Grad{p} +(\bv{v}_\f-\bv{v}_\s)=\bv{0},
\end{equation}
which, under the physically reasonable assumption that $\bv{a}_{\f} - \bv{b}_{\f}$ and $\Grad{p}$
are bounded fields over $B_{t}$, fulfills the assumed limit-case condition that $\bv{v}_\f\to\bv{v}_\s$ as $\phi_\f\to \np{0}$.
\begin{remark}[On the boundedness of physical quantities]
    From a physical perspective, it is indeed reasonable to assume that $\bv{a}_{\f} - \bv{b}_{\f}$ and $\Grad{p}$
    be bounded fields over $B_{t}$. In general, $\bv{b}_{\f}$ is a prescribed field and therefore its boundedness can be taken for granted. However, the boundedness of the other two quantities 
    ($\Grad{p}$ in particular) 
    cannot be taken for granted from the perspective of the theory of partial differential equations depending on topological considerations and the regularity of the data of the underlying boundary value problem.
\end{remark}

The equation of motion for the mixture, that is, Eq.~\eqref{equ:eom_mixture}, is not affected by the limit case condition, since $\phi_\f$ does not directly contribute to this equation. Indeed, as $\phi_\f\to \np{0}$, $\rho_\f$ approaches zero, and the equation of motion for the solid component is recovered from Eq.~\eqref{equ:eom_mixture}, as expected.
The equation arising from the balance of mass and incompressibility, namely Eq.~\eqref{equ:mass_balance_mixture}, remains meaningful in the limit case. In fact, recalling the saturation condition $\phi_{\f}+\phi_{\s} = 1$, we have that $\phi_{\s} \bv{v}_{\s} + \phi_{\f}\bv{v}_{\f} = \bv{v}_{\s} + \bv{v}_{\flt}$. Hence, recalling that $\|\bv{v}_{\flt}\|\to \np{0}$ as $\phi_{\f}\to \np{0}$, for the limit behavior of Eq.~\eqref{equ:mass_balance_mixture} we have
\begin{equation}
    \lim_{\phi_\f\to 0} \Div{(\phi_\s\bv{v}_\s+\phi_\f\bv{v}_\f)} = 0
    \quad\Rightarrow\quad \Div{\bv{v}_\s} = 0,
\end{equation}
which is the expression of the incompressibility of the pure solid phase (in rate form).

There is no unique way to satisfy Eq.~\eqref{equ: permeability assumption}. Possible practicable approaches can be derived by considering the experimental literature (e.g., see \cite{Yuan2023PorosityPermeab}). In addition to experimental considerations, which require to be adapted to specific physical systems, considerations can be included pertaining to practical implementations within a numerical framework.
In order to meet the condition stated in Eq.~\eqref{equ:eta_cond}, and for the purpose of illustrating some practical examples, here we propose the following definition for $\kappa_\s(\phi_\f)$
\begin{equation}\label{equ:definition_ks}
	\kappa_\s(\phi_\f) \coloneqq 
	\begin{cases} 
		\overline{\kappa}_\s\left(4-3\dfrac{\phi_\f}{\phi_{\f_\mathrm{c}}}\right) \left(\dfrac{\phi_\f}{\phi_{\f_\mathrm{c}}}\right)^3  & \phi_\f< \phi_{\f_\mathrm{c}}, \\
		\overline{\kappa}_\s & \phi_\f\ge\phi_{\f_\mathrm{c}},
	\end{cases}
\end{equation}
where $\overline{\kappa}_\s$ is the saturation value of permeability and $\phi_{\f_\mathrm{c}}$ is a critical porosity value.

\begin{figure}[ht]
	\centering
	\includegraphics[width=0.5\textwidth]{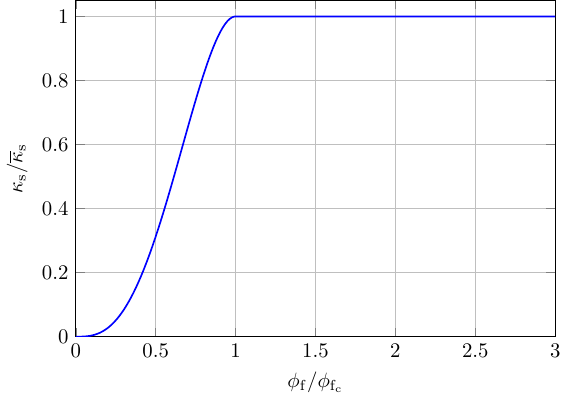}
	\caption{Plot of $\kappa_\s/\overline{\kappa}_\s$ as a function of $\phi_{\f}/\phi_{\f_\mathrm{c}}$ to illustrate the behavior of the permeability $\kappa_\s$ of the solid component as defined in Eq.~\eqref{equ:definition_ks}. The permeability remains constant when $\phi_{\f} > \phi_{\f_\mathrm{c}}$.}
	\label{fig:kappa_phi}
\end{figure}

The behavior of $\kappa_\s(\phi_\f)/\overline{\kappa}_\s$ for $0\leq \phi_\f\leq 1$ is shown in Fig.~\ref{fig:kappa_phi}, for specific choice of parameter $\phi_{\f_\mathrm{c}}$. It is worth mentioning that definition~\eqref{equ:definition_ks} of $\kappa_\s(\phi_\f)$ conforms to some of the experimental data available in the literature (cf.\ \cite{Yuan2023PorosityPermeab}); clearly, other functional forms of solid permeability can be employed.

\section{Weak formulations and verification}\label{sec: weak form}
In this section, we present the weak form that we choose for the problem's governing equations. The weak form in question is then implemented in a \ac{FEM} for the solution of the examples presented in this paper. Overall, the \ac{FEM} we implement is an \ac{ALE} scheme (cf.\ \cite{Fernandez2009The-Derivation-0}) where the computational domain is chosen as the initial configuration of the solid phase, the latter having been chosen to coincide with $B_{\s}$.

\subsection{Formulation}
We begin by rewriting the equations as a single system:  
\begin{align}
\label{equ:strong_form_ref, fluid}
    \frac{\phi_\f \rho^*_\f}{\eta(\phi_\f)} (\bv{a}_\f-\bv{b}_\f)+\dfrac{\phi_\f}{\eta(\phi_\f)} \Grad{p} +(\bv{v}_\f-\bv{v}_\s)&=\bv{0} & \text{in $B_{t}$},        
    \\
    \label{equ:strong_form_ref, mixture}
    \rho^*_\f(J_\s-\phi_{R_\s})(\bv{a}_\f-\bv{b}_\f)^{\circd{s}} + \rho^*_\s\phi_{R_\s}(\bv{a}_\s-\bv{b}_\s)^{\circd{s}} - \Div{\ts{P}} &= \bv{0} & \text{in $B_{\s}$},       
    \\
    \label{equ:strong_form_ref, continuity}
    \Div{\left(
    \phi_{\s} \bv{v}_{\s} +
    \phi_{\f} \bv{v}_{\f}
    \right)}&=0 & \text{in $B_{t}$},
\end{align}
where $\eta(\phi_\f)$ is given in Eq.~\eqref{equ:eta_def}, and where
\begin{equation}
\ts{P} = -J_\s \, p^{\circd{s}}\, 
\inversetranspose{\ts{F}}_\s + \ts{P}^\mathrm{e}_\s
\end{equation}
such that
\begin{equation}
\ts{P}^\mathrm{e}_\s = \phi_{R_\s}\ts{F}_\s\ts{S}^{\mathrm{e},*}_\s
\quad\text{and}\quad
\ts{S}^{\mathrm{e},*}_\s = 2 \dfrac{\partial \hat{\Psi}_\s}{\partial \ts{C}_\s}.
\end{equation}
Here, $\ts{P}$ is the first Piola-Kirchhoff stress tensor for the mixture, $\ts{P}^\mathrm{e}_\s$ is the first Piola-Kirchhoff elastic stress tensor for the solid phase, and $\ts{S}^{\mathrm{e},*}_\s$ is the second Piola-Kirchhoff elastic stress tensor for the pure solid phase.

The strong form of the equations will now be reformulated in a weak form after we introduce appropriate functional spaces for the primary unknown fields. From a physics's perspective, we choose the problem's primary unknowns to be the solid's displacement $\bv{u}_{\s}$, the fluid's velocity $\bv{v}_{\f}$, and the pore pressure $p$. However, we will solve the governing equations by adopting an \ac{ALE} scheme according to which the equations are pulled back to $B_{\s}$. Therefore, from an implementation perspective, the primary unknowns are $\bv{u}_{\s}$, $\bv{v}_{\f}^{\circd{s}}$, and $p^{\circd{s}}$, where $p^{\circd{s}}$ and $\bv{v}_{\f}^{\circd{s}}$ are viewed as primitive quantities, whereas $\bv{v}_{\f}$ and $p$ are images under $\bv{\chi}_{\s}$ of $p^{\circd{s}}$ and $\bv{v}_{\f}^{\circd{s}}$. 

To facilitate the expression of the boundary conditions, let $\Gamma_{\s}$ denote the boundary of $B_{\s}$, i.e., $\partial B_\s$ in Fig.~\ref{fig:configurations}. With $i=1,\ldots,d$, we then consider the following partitions:
\begin{alignat}{2}
\label{equ: Gamma_s partition u}
\Gamma_{\s} &= \Gamma_{\s_{i}}^{D\bv{u}_{\s}} \cup \Gamma_{\s_{i}}^{N\bv{u}_{\s}},
&&\quad\Gamma_{\s_{i}}^{D\bv{u}_{\s}} \cap \Gamma_{\s_{i}}^{N\bv{u}_{\s}} = \emptyset,
\\
\label{equ: Gamma_s partition p}
\Gamma_{\s} &= \Gamma_{\s}^{Dp} \cup \Gamma_{\s}^{Np},
&&\quad\Gamma_{\s}^{Dp} \cap \Gamma_{\s}^{Np} = \emptyset,
\end{alignat}
where the superscripts `$D$' and `$N$' stand for `Dirichlet' and `Neumann', respectively. Then, with a slight abuse of notation, let $\bv{u}_\s \in \mathcal{V}^{\bv{u}_\s}$, $\bv{v}_\f \in \mathcal{V}^{\bv{v}_\f}$, and $p \in \mathcal{V}^p$, with
\begin{align}
\label{equ: us functional space}
\mathcal{V}^{\bv{u}_\s} & \coloneqq \left\{\bv{u}_\s \in \left[L^2\left(B_\s\right)\right]^d \mid
\Grad{\bv{u}_\s} \in \left[L^{\infty}\left(B_\s\right)\right]^{d \times d},\text{~$u_{\s_{i}} = u_{\g_{i}}$ on $\Gamma_{\s_{i}}^{D_{\bv{u}_{\s}}}$} \right\},
\\
\label{equ: vf functional space}
\mathcal{V}^{\bv{v}_\f} & \coloneqq \left\{\bv{v}_\f \in \left[L^2\left(B_t\right)\right]^d \mid
\Grad
{\bv{v}_\f} \in \left[L^2\left(B_t\right)\right]^{d \times d}\right\},
\\
\label{equ: p functional space}
\mathcal{V}^p & \coloneqq \left\{p \in \left[L^2\left(B_t\right)\right]^d \mid
\Grad
{p}
\in \left[L^2\left(B_t\right)\right]^{d \times d},
\text{~$p=p_{\g}$ on $\bigl( \Gamma_\s^{D_p}\bigr)^{\circd{e}}$}
\right\},
\end{align}
where $u_{\g_{i}}$ is the Dirichlet data for the $i$-th component of $\bv{u}_{\s}$, $p_{\g}$ is the Dirichlet data for the pore pressure field, and where $\bigl(\Gamma_\s^{D_{p}}\bigr)^{\circd{e}}$ denotes the image of $\Gamma_\s^{D_{p}}$ under $\bv{\chi}_{\s}$. We also define 
\begin{align}
\label{equ: test of us functional space}
    \mathcal{V}_0^{\bv{u}_\s} &\coloneqq \left\{\bv{u}_\s \in \left[L^2\left(B_\s\right)\right]^d \mid \Grad{\bv{u}_\s} \in \left[L^{\infty}\left(B_\s\right)\right]^{d \times d},\text{~$u_{\s_{i}} = \np{0}$ on $\Gamma_{\s_{i}}^{D_{\bv{u}_{\s}}}$} \right\},
    \\
    \mathcal{V}_0^p &\coloneqq \left\{p \in \left[L^2\left(B_t\right)\right]^d \mid \Grad{p} \in \left[L^2\left(B_t\right)\right]^{d \times d},\text{~$p=\np{0}$ on $\bigl(\Gamma_\s^{D_p}\bigr)^{\circd{e}}$}
    \right\}.
\end{align}
\begin{remark}[On the regularity of $\bv{u}_{\s}$]
The overall problem will be solved via an \ac{ALE} scheme that remaps all equations over $B_{\s}$. Hence, the regularity of $\bv{u}_{\s}$ indicated in Eq.~\eqref{equ: us functional space} is needed to guarantee the assumed regularity of the remaining fields.
\end{remark}

\begin{remark}[On the regularity of $p$]
The choice of $\mathcal{V}^{p}$ is motivated by the way in which we weakly impose Eq.~\eqref{equ:strong_form_ref, continuity}. Specifically, after multiplying the latter by an appropriate test function, we will integrate by parts the resulting equation. In turn, this approach is motivated by a desire to avoid dealing with gradients of the mixture's volume fractions.
\end{remark}

Depending on the context, by the notation $(\cdot,\cdot)_{\square}$ we will denote the scalar product over the domain `$\square$'. Furthermore,
to avoid proliferation of symbols, we denote the test functions for a field by placing a tilde over the symbol of the field in question. 

As the weak form of the equation of motion for the mixture in the reference configuration, we choose the following statement, which we impose for all $\tilde{\bv{u}}_\s \in \mathcal{V}^{\bv{u}_\s}_0$:
\begin{equation}\label{equ:weakform_mixture_ref}
	\begin{multlined}[b]
		\Bigl(\Tilde{\bv{u}}_\s,\rho^*_\f(J_\s-\phi_{R_\s})(\bv{a}_\f-\bv{b}_\f)^{\circd{s}}\Bigl)_{B_{\s}}+
		\Bigl(\Tilde{\bv{u}}_\s,\rho^*_\s\phi_{R_\s}(\bv{a}_\s-\bv{b}_\s)^{\circd{s}}\Bigl)_{B_{\s}}
            \\
            +\Bigl(\Tilde{\bv{u}}_\s,J_\s\inversetranspose{\ts{F}_\s}\Grad{(p^{\circd{s}})}\Bigl)_{B_{\s}}
            +
            \Bigl(\Grad{\Tilde{\bv{u}}}_\s,\ts{P}^\mathrm{e}_\s\Bigl)_{B_{\s}}
            -
		\sum_{i=1}^{d}
            \Bigl(\Tilde{u}_{\s_{i}}, \sigma_{i}
            \Bigl)_{\Gamma_{\s_{i}}^{N_{\bv{u}}}} = 0,
	\end{multlined}
\end{equation}
where, letting $s_{\g_{i}}$ be the prescribed $i$-th component of the traction acting on the boundary of the current configuration of the mixture, and recalling that the outward unit normal on $\partial B_{t}$ is $\bv{n} = \bigl(\inversetranspose{\ts{F}}_{\s}\bv{n}_{\s}/\|\inversetranspose{\ts{F}}_{\s}\bv{n}_{\s}\|\bigr)^{\circd{e}}$,
\begin{equation}
\label{equ: sigma def}
\sigma_{i} \coloneqq \biggl(
s_{\g_{i}}^{\circd{s}} + p^{\circd{s}} \frac{\bv{e}_{i} \cdot \inversetranspose{\ts{F}}_{\s}\bv{n}_{\s}}{\|\inversetranspose{\ts{F}}_{\s}\bv{n}_{\s}\|}
\biggr) J_{\s} \|\inversetranspose{\ts{F}}_{\s}\bv{n}_{\s}\|.
\end{equation}
Differently from what is typically done in mixed \ac{FEM} for fluids problems, the weak form in Eq.~\eqref{equ:weakform_mixture_ref} does not manage the pressure gradient terms in the strong form of the equation through the lens of the divergence theorem.

We choose as the weak form of the equation of motion for the fluid the following statement, that we impose for all $\tilde{\bv{v}}_\f \in \mathcal{V}^{\bv{v}_\f}$:
\begin{equation}\label{equ:weakform_fluid}
    \Bigl(\Tilde{\bv{v}}_\f,\frac{\phi_\f \rho^*_\f}{\eta(\phi_\f)}(\bv{a}_\f-\bv{b}_\f)
    +
    \frac{\phi_\f}{\eta(\phi_\f)}\Grad{p}
    +
    \bv{v}_\f-\bv{v}_\s\Bigl)_{B_{t}} = 0.
\end{equation}
Note that in Eq.~\eqref{equ:weakform_fluid} integration by parts is not applied to the pressure term.

Finally, we consider the weak form of the continuity equation. Before doing so, it is important to remark that in poroelasticity problems in which the porous phase and the fluid interaction is described by the last term on the left-hand side of Eq.~\eqref{equ:eom_fluid}, there is a boundary condition that can be enforced limited to the normal component of the filtration velocity. In the most traditional form of the Darcy flow problem where the solid phase is assumed to be stationary, this condition corresponds to prescribing the normal component of the fluid's velocity on the boundary of the domain (cf.\ \cite{Masud2007A-Stabilized-Mixed-0}). Referring to Eq.~\eqref{eq: vflt def} and the boundary partitions in Eqs.~\eqref{equ: Gamma_s partition u} and~\eqref{equ: Gamma_s partition p}, let $v_{\flt_{\g}}$ denote the prescribed value of $\bv{v}_{\flt}\cdot\bv{n}$ on $(\Gamma_{\s}^{N_{p}})^{\circd{e}}$, then we weakly enforce Eq.~\eqref{equ:strong_form_ref, continuity} by requiring that, for all $\tilde{p}\in\mathcal{V}^{p}_{0}$, we have
\begin{equation}\label{equ:weakform_continuity}
    \Bigl(\Tilde{p}, \bv{v}_{\s} \cdot \bv{n} + v_{\flt_{\g}} \Bigl)_{(\Gamma_\s^{N_p})^{\circd{e}}}-
    \Bigl(\Grad{\Tilde{p}},\phi_\s\bv{v}_\s+\phi_\f\bv{v}_\f \Bigl)_{B_{t}} = 0.
\end{equation}

In conclusion, we determine the motion of the poroelastic mixture under consideration by solving the following
\begin{problem}
    Find $\bv{u}_{\s}$, $\bv{v}_{\f}^{\circd{s}}$, and $p^{\circd{\s}}$ in $\mathcal{V}^{\bv{u}_{\s}}$, $\mathcal{V}_{\s}^{\bv{v}_{\f}}$, and $\mathcal{V}_{\s}^{p}$, respectively, such that for all $\tilde{\bv{u}}_{\s}$, $\tilde{\bv{v}}_{\f}$, and $\tilde{p}$ in $\mathcal{V}_{0}^{\bv{u}_{\s}}$, $\mathcal{V}_{\s}^{\bv{v}_{\f}}$, and $\mathcal{V}_{0,\s}^{p}$, respectively, 
Eqs.~\eqref{equ:weakform_mixture_ref}, \eqref{equ:weakform_fluid}, and~\eqref{equ:weakform_continuity} are satisfied, where $\mathcal{V}_{\s}^{\bv{v}_{\f}} = \left( \mathcal{V}^{\bv{v}_{\f}} \right)^{\circd{s}}$, $\mathcal{V}_{\s}^{p} = \left( \mathcal{V}^{p} \right)^{\circd{s}}$ and $\mathcal{V}_{0,\s}^{p} = \left( \mathcal{V}_{0}^{p} \right)^{\circd{s}}$.
\end{problem}

\begin{remark}[Consistency with the strong form]
    The fact that the weak problem we have formulated is equivalent to that presented in strong form can be shown via standard methods relying on integration by parts and therefore is omitted here.
\end{remark}

\begin{remark}[On the unusual presentation of the weak forms]
    The presentation of the weak problem we solve relies on a slight abuse of notation in that we do not present equations all of which are pulled back to the chosen solution domain, namely $B_{\s}$. The reason for this choice is twofold. First, the equation of motion for the fluid along with the continuity equation are easier to read and to relate to the strong form of the equations when presented in Eulerian form (i.e., using functions with domain $B_{t}$). Second, from a computer implementation viewpoint, the problem was solved using \ac{CMPH}~\cite{COMSOLCite}. The latter can be used as a finite element library supporting \ac{ALE} schemes whereby operators implemented in Eulerian form can be evaluated while taking into account a given \ac{ALE} map. This is to say that we presented the equations in the way we actually implemented them in \ac{CMPH}.
\end{remark}

\subsection{Verification by method of manufactured solutions}\label{subsec:mms}

Here we use the method of manufactured solutions \cite{Salari2000CodeVerificatio} to verify the correctness of our computer implementation, and to empirically verify that the proposed formulation can indeed deliver results under conditions of zero porosity. This analysis also includes the empirical assessment of the convergence rates of our numerical approach.

We selected finite element spaces within a 2D axisymmetric domain, whose reference configuration, denoted by $\mathrm{A}$, has a radius of $R = \np[mm]{5}$ and a height of $L = \np[mm]{10}$. The deformed configuration is denoted by $\mathrm{A}(t)$.
The manufactured solution consists of the following fields:
\begin{align}
\label{eq: MMS p}
    p &= p_o \sin(\pi t/(2 t_o)) \bigl(r/L\bigr)^3 \bigl(z/L\bigr)^3 & & \text{in $\mathrm{A}(t)$},\\
\label{eq: MMS us r}
     u_{\s_{R}} &= u_o \bigl[1-\cos(\pi t/(2 t_o))\bigr] \bigl(R/L\bigr)^3 & & \text{in $\mathrm{A}$},\\
\label{eq: MMS us Z}
     u_{\s_{Z}} &= u_o \bigl[1-\cos(\pi t/(2 t_o))\bigr] \bigl(Z/L\bigr)^3 & & \text{in $\mathrm{A}$},\\
\label{eq: MMS vf}
     \bv{v}_\f  &= (\partial_t \bv{u}_\s)^{\circd{e}} - (\phi_\f/\eta) 
     \Grad{p}
     & & \text{in  $\mathrm{A}(t)$},
\end{align}
where $p_o=\np[Pa]{100}$, $u_o=\np[mm]{0.1}$, and $t_o=\np[s]{1}$. Moreover, $(R,Z)$ and $(r,z)$ are the coordinates in the reference and deformed configurations, respectively.
We considered two cases: one with $\phi_{\f} = \np{0.5}$ and the other with $\phi_{\f} = \np{0}$ with all other parameters as in Table~\ref{tab:MMS Example params}.
\begin{table}[hbt]
\centering
\caption{Parameter values used in the convergence rate study and the case study.}
\label{tab:MMS Example params}
\begin{tabular}{lll}
	\hline Parameter & Description & Value \\
	\hline
        $\rho^*_\s$ & Solid true density & $\np[kg/m^3]{1e3}$ \\
        $\rho^*_\f$ & Fluid true density & $\np[kg/m^3]{1e3}$ \\
        $\mu_\f$ & Fluid viscosity & $\np[Pa\cdot s]{1e{-3}}$ \\
	$\overline{\kappa}_\s$ & Solid Saturated permeability &  $\np[m^2]{1e{-9}}$ \\
	$\phi_{\f_\mathrm{c}}$ & Critical porosity& $\np{0.1}$\\
        $\mu_\s$ & Solid shear modulus & $\np[kPa]{5}$\\
	\hline
\end{tabular}
\end{table}
Referring to Eq.~\eqref{eq: MMS vf}, we note that our chosen manufactured solution is such that, when $\phi_{\f} = \np{0}$, the velocity of the fluid is the same as the velocity of the solid.

The initial conditions correspond to the manufactured solution evaluated at $t=\np{0}$. For the solid phase, boundary conditions are imposed in the form of prescribed displacement along the entire external boundary, except the top edge where the elastic traction obtained form the manufactured solution is applied.
For the fluid phase, the normal component of the filtration velocity, as determined from the manufactured solution, is enforced along the external boundaries. 
Along the axis of symmetry, the radial component of the displacement field is set to zero, while the horizontal component is prescribed by the manufactured solution; similarly, the normal component of the filtration velocity is imposed based on the manufactured solution.
The formulation is numerically implemented using \ac{CMPH}~\cite{COMSOLCite}, which serves as a \ac{FEM} programming environment, and solved adopting a monolithic solver with Newton's method whose damping factor is determined automatically. First-order Lagrange polynomial interpolation over triangles is used for all fields except for the solid displacement. For the latter, we employ a MINI element ($P_{1}$ element enriched with a cubic bubble) \cite{Arnold1984AStableFiniteEl}. A mesh composed of identical triangular elements is utilized, and five successive uniform refinement iterations are performed. The initial characteristic mesh size is $h_e=\np[mm]{2.5}$.
For time integration, second-order \ac{BDF} is used with a constant time step of $\np[s]{5e-4}$, and the simulation is carried out until a final time of $\np[s]{5e-2}$.

\begin{remark}[On finite element spaces]
 The numerical implementation of our weak formulation produces a mixed \ac{FEM} (cf.\ \cite{Brezzi1991Mixed-and-Hybrid-0}). As such, one must confront the problem of selecting appropriate stable finite element spaces. A detailed discussion of the issue at hand is outside the scope of this paper and we simply point to references \cite{Masud2007A-Stabilized-Mixed-0,Rodrigo2018New-Stabilized-,Barnafi2021Mathematical-An} for context. This said, the use of the method of manufactured solutions allowed us to empirically confirm that our choice of finite element spaces appears to be stable. To some extent, this finding is not surprising in that, in the vanishing porosity limit, the solid's response is that of an incompressible hyperelastic system and our selection of finite element spaces is known to be stable for these problems (cf.\ \cite{Karabelas2020Versatile-Stabi,Karabelas2022An-Accurate-Rob}).
\end{remark}

\begin{figure}[phtb]
	\begin{subfigure}[b]{0.4\textwidth}
		\centering
		\includegraphics[width=0.8\textwidth]{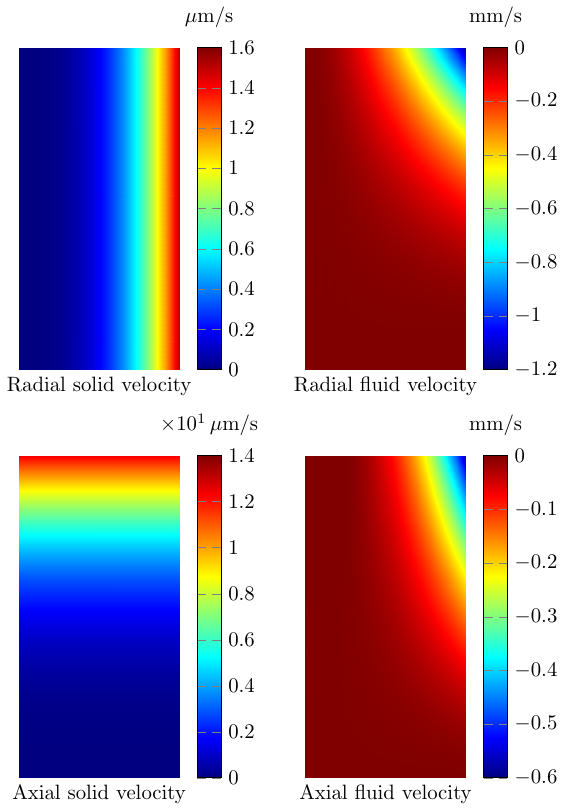}
		\caption{Comparison of fluid and solid velocity for the case $\phi_{f} = \np{0.5}$ at time $t=\np[s]{5e-2}$.}
		\label{fig:mms phif=0.5}
	\end{subfigure}
	\hfill
	\begin{subfigure}[b]{0.55\textwidth}
		\centering
		\includegraphics[width=0.95\textwidth]{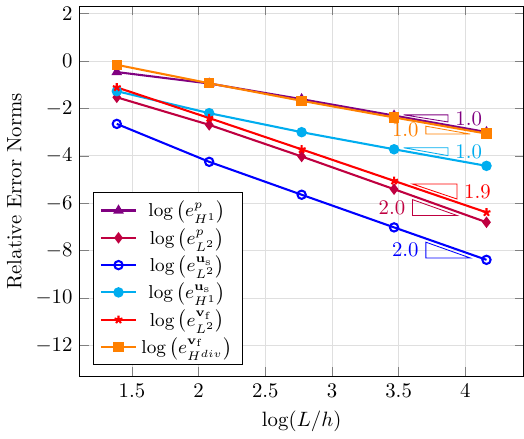}
		\caption{Relative error norms for the case $\phi_{f} = \np{0.5}$ at time $t=\np[s]{5e-2}$.}
		\label{fig:conv_rates phif=0.5}
	\end{subfigure}

    \begin{subfigure}[b]{0.4\textwidth}
		\centering
		\includegraphics[width=0.8\textwidth]{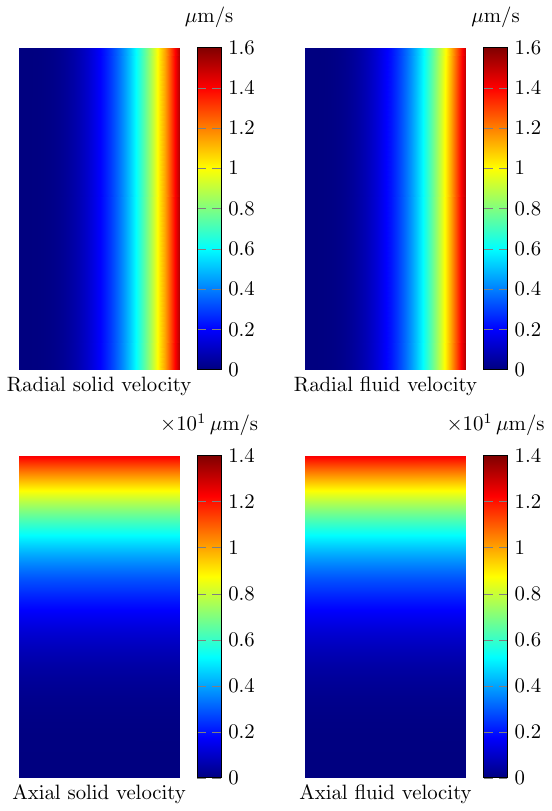}
		\caption{Comparison of fluid and solid velocity for the case $\phi_{f} = \np{0}$ at time $t=\np[s]{5e-2}$.}
		\label{fig:mms phif=0}
	\end{subfigure}
	\hfill
	\begin{subfigure}[b]{0.55\textwidth}
		\centering
		\includegraphics[width=0.95\textwidth]{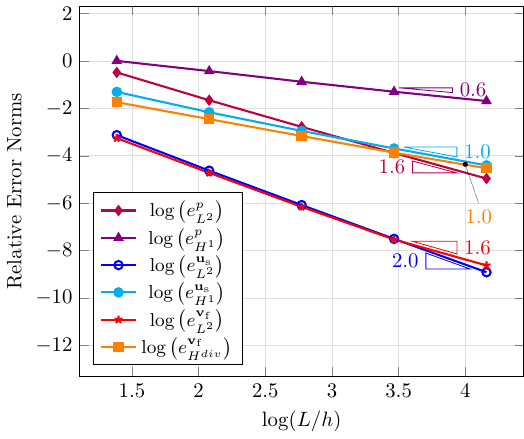}
		\caption{Relative error norms for zero porosity case at time $t=\np[s]{5e-2}$.}
		\label{fig:conv_rates phif=0}
	\end{subfigure}
	\caption{Model verification via the method of manufactured solutions, with $\phi_{f} = \np{0.5}$ in the first row and $\phi_{f} = \np{0}$ in the second row. \subref{fig:mms phif=0.5} and~\subref{fig:mms phif=0} show radial and axial components of the velocity field for the solid (left) and fluid phase (right) for $\phi_{f} = \np{0.5}$ and $\phi_{f} = \np{0}$, respectively. As expected, both components of the fluid velocity converge to those of the solid phase under the limit case condition.
    \subref{fig:conv_rates phif=0.5} and~\subref{fig:conv_rates phif=0} represent convergence rates for relative $\triangle$-norm (all fields' $L^2$-norm, pressure's and solid displacement's $H^1$-semi-norms, and fluid velocity's $H^{div}$-semi-norm) 
    of the error $e^\square_\triangle=\|\square-\blacksquare\|_\triangle/\|\square\|_\triangle$ between the numerical solution $\blacksquare$ and exact solution $\square$ at time $t = \np[s]{5e-2}$ for two cases of $\phi_{f} = \np{0.5}$ and $\phi_{f} = \np{0}$, respectively, obtained using P1 interpolation for all fields and a cubic bubble augmentation for the solid displacement (i.e., a MINI element \cite{Arnold1984AStableFiniteEl}).%
    }
	\label{fig:mms_conv_rates}
\end{figure}

Figures~\ref{fig:mms_conv_rates}\subref{fig:mms phif=0.5} and~\ref{fig:mms_conv_rates}\subref{fig:mms phif=0} show color maps of the radial and axial velocity components of the solid and fluid for $\phi_{\f} = \np{0.5}$ and $\phi_{\f} = \np{0}$, respectively, as obtained from our calculations using the most refined mesh.
As seen in Fig.~\ref{fig:mms_conv_rates}\subref{fig:mms phif=0}, the fluid velocity converges to that of the solid in the limit case, as discussed in Section~\ref{sec: reformulation}.
Figures~\ref{fig:mms_conv_rates}\subref{fig:conv_rates phif=0.5} and~\ref{fig:mms_conv_rates}\subref{fig:conv_rates phif=0} present the convergence rates for $\phi_{\f} = \np{0.5}$ and $\phi_{\f} = \np{0}$, respectively.
As shown in Fig.~\ref{fig:mms_conv_rates}\subref{fig:conv_rates phif=0}, the convergence rates for the solid displacement and pressure error norms in the limit case are similar to those observed in a pure solid problem (cf.\ \cite{Masud2013A-Framework-for}).

The results in Fig.~\ref{fig:mms_conv_rates} show that our numerical implementation of our proposed model is correct and that it is indeed capable of successfully handling even a case with porosity equal to zero. We note that the convergence rates for the fluid velocity are as one would normally expect them for a flow problem like, say, Stokes's, with P1 interpolation for the fluid velocity (cf.\ \cite{Masud2007A-Stabilized-Mixed-0}).
 
\subsection{Case study: local compression of a porous cylinder}\label{subsec:benchmark}

\begin{figure}[phbt]
	\begin{minipage}{0.5\linewidth}
		\begin{subfigure}[b]{\textwidth}
			\centering
			\includegraphics[width=\textwidth]{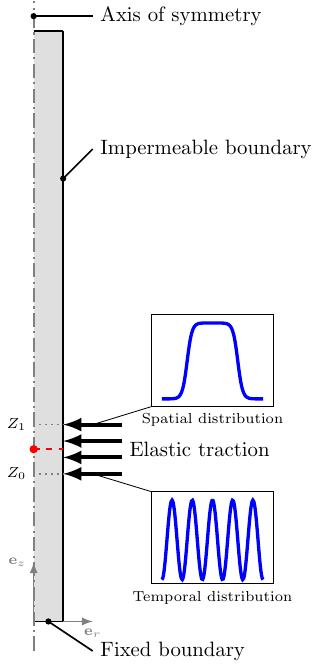}
			\caption{Problem setup}
			\label{fig:ExampleProblemSetup}
		\end{subfigure}
	\end{minipage}
	\hfill
	\begin{minipage}{0.5\linewidth}
		\begin{subfigure}[b]{\textwidth}
			\centering
			\includegraphics[width=0.64\textwidth]{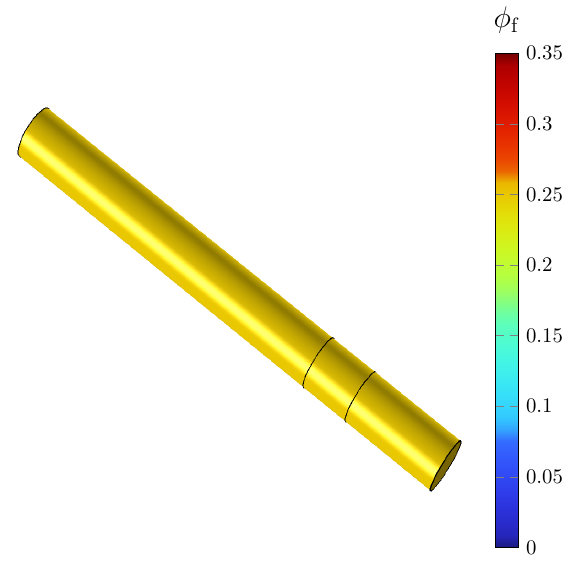}
			\caption{$\phi_\f$ at $t=\np[s]{0}$: homogeneous initial fluid volume fraction.}
			\label{fig:ExampleProblemInitial}
		\end{subfigure}
		\hfill
		\begin{subfigure}[b]{\textwidth}
			\centering
			\includegraphics[width=0.64\textwidth]{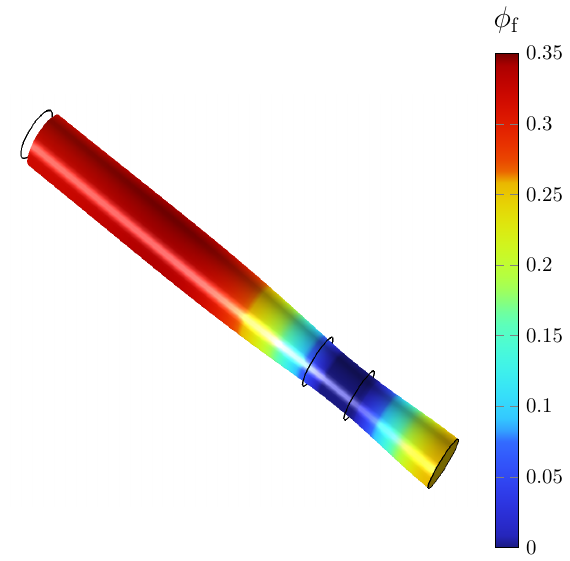}
			\caption{$\phi_\f$ at $t=\np[s]{2}$: the squeezing zone shaded in dark blue is free of fluid after five squeezing periods.}
			\label{fig:ExampleProblemFinal}
		\end{subfigure}
        \hfill
		\begin{subfigure}[b]{\textwidth}
			\centering
			\includegraphics[width=0.8\textwidth]{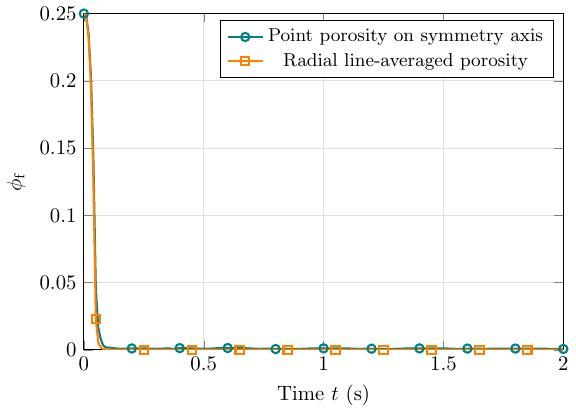}
			\caption{Porosity as a function of time.}
			\label{fig:ExampleProblemPorosityTime}
		\end{subfigure}
	\end{minipage}
	\caption{Squeezing test for a full cylinder modeled as a 2D axisymmetric domain. \subref{fig:ExampleProblemSetup} Problem setup. \subref{fig:ExampleProblemInitial} Initial and \subref{fig:ExampleProblemFinal} deformed configuration. \subref{fig:ExampleProblemPorosityTime} Evolution of porosity over time at a point located at the center of the squeezing zone along the symmetry axis, i.e. at the coordinate $\bigl(\np{0},\left(Z_0+Z_1\right)/2\bigr)$, and the average porosity over the radial line passing through the same point, as shown in Fig. \subref{fig:ExampleProblemSetup} in red color. In the initial state, the fluid is homogeneously distributed. Under the applied traction, the fluid redistributes, resulting in fluid-free regions and regions where fluid and solid coexist.}
	\label{fig:ExampleProblem}
\end{figure}

To further investigate the performance of the proposed model under local extreme conditions, we simulate a benchmark problem involving a cylinder that is locally compressed. As the reference configuration (cf.\ Fig.~\ref{fig:ExampleProblemSetup}), we consider a 2D axisymmetric geometry with a radius of $R = \np[mm]{0.5}$ and a height of $L = \np[mm]{10}$. 
The cylindrical domain is subject to radial squeezing due to a time-harmonic traction distribution purely in the (current) radial direction applied to the region located between $Z_0=\np[mm]{2}$ and $Z_1=\np[mm]{3}$ from the bottom boundary, as defined by the following equation and illustrated in Fig.~\ref{fig:ExampleProblem}\subref{fig:ExampleProblemSetup},
\begin{equation*}
\bv{p}_\g=J_{\s} \|\inversetranspose{\ts{F}}_{\s}\bv{n}_{\s}\| \bv{s}_\g=-p_o \text{rect}\left(\frac{Z-Z_0}{Z_1-Z_0}\right) \left(1 - \cos\left(2\pi\frac{ t}{t_o} \right)\right)\bv{e}_r,
\end{equation*}
where $p_o=\np[Pa]{100}$ and $t_o=\np[s]{0.4}$. Moreover, $\text{rect}(s)$ is a rectangular function defined to be twice continuously differentiable and to attain unity within the interval $\left[0.2, 0.8\right]$ of its domain $\left[0, 1\right]$, for which we employed the implementation available in \ac{CMPH}~\cite{COMSOLCite}.
The bottom boundary is fixed, whereas the top and lateral edges, except the squeezing zone, are traction-free. Zero pressure is prescribed at the top and bottom boundaries, while the lateral surface is impermeable (i.e., $\bv{v}_\flt\cdot\bv{n}=\np{0}$). 
The radial displacement and the normal component of the filtration velocity are constrained to zero along the axis of symmetry.
Furthermore, quiescent initial conditions are imposed. The material properties employed in this study are reported in Table~\ref{tab:MMS Example params}. A uniform mesh made up of identical triangular elements is used, with an element size of $\np[mm]{0.1}$. Time integration is carried out using second-order \ac{BDF} with a constant time step of $\np[s]{1e-3}$, and the simulation runs until a final time of $\np[s]{2}$. All other settings, including the element type and solver configurations, are consistent with those adopted for the verification test in Section \ref{subsec:mms}.

\begin{figure}[htb]
\centering
\begin{subfigure}[b]{0.3\textwidth}
    \centering
    \includegraphics[height=2.2in]{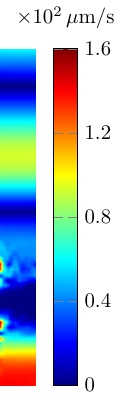}
    \caption{}
    \label{fig:ExampleProblem Filtration Velocity Norm}
\end{subfigure}
\begin{subfigure}[b]{0.3\textwidth}
    \centering
    \includegraphics[height=2.2in]{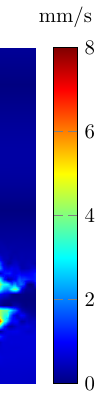}
    \caption{}
    \label{fig:ExampleProblem Relative Velocity Norm}
\end{subfigure}
\begin{subfigure}[b]{0.3\textwidth}
    \centering
    \includegraphics[height=2.2in]{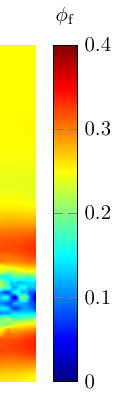}
    \caption{}
    \label{fig:ExampleProblem Traditional Formulation}
\end{subfigure}
\caption{\subref{fig:ExampleProblem Filtration Velocity Norm} Magnitude of the filtration velocity $\|\bv{v}_\flt\|$ at $t=\np[s]{2}$. It can be observed that the filtration velocity approaches zero wherever the porosity vanishes.
\subref{fig:ExampleProblem Relative Velocity Norm} Magnitude of the relative velocity $\|\bv{v}_\f - \bv{v}_\s\|$ at $t = \np[s]{2}$. In regions where the porosity vanishes, the fluid and solid velocities tend to converge. It is worth noting that in regions far from the squeezing zone, both the solid and fluid velocities are zero.
\subref{fig:ExampleProblem Traditional Formulation} The porosity $\phi_\f$ obtained using the traditional poroelastic formulation. The simulation stops computing in the middle ($t=\np[s]{5e-2}$) because the traditional formulation is numerically unstable as the porosity becomes very small ($\np{5e-4}$).}
\label{fig:ExampleProblem Analysis}
\end{figure}

The aim of this study is to evaluate the model's ability to accurately represent regions of locally vanishing porosity.
When the traction force is applied, the fluid is driven out of the porous medium. As a result, some areas within the porous medium become devoid of fluid, while other areas retain both solid and fluid phases simultaneously, as shown in Fig.~\ref{fig:ExampleProblem}\subref{fig:ExampleProblemFinal}.
Moreover, as demonstrated in Fig.~\ref{fig:ExampleProblem}\subref{fig:ExampleProblemPorosityTime}, both the porosity at the center of the squeezing zone along the symmetry axis and the average porosity along a radial line passing through the middle of the squeezing zone (see Fig.~\ref{fig:ExampleProblem}\subref{fig:ExampleProblemSetup}) decrease rapidly during the initial phase of the applied elastic traction.
Figures~\ref{fig:ExampleProblem Analysis}\subref{fig:ExampleProblem Filtration Velocity Norm} and~\subref{fig:ExampleProblem Relative Velocity Norm} show the norm of the filtration velocity and the norm of the relative velocity of the fluid with respect to the solid. As the porosity approaches zero, both the filtration velocity and the relative velocity (though at a slower rate) tend to zero, confirming the assumption that the fluid velocity converges to the solid velocity.
Therefore, the model effectively simulates the behavior of porous media under extreme conditions, where some regions become devoid of fluid. These findings add to the ones from Section \ref{subsec:mms}, where it was shown that the proposed finite element spaces demonstrate stability and acceptable convergence rate.

It is worth emphasizing that the conventional formulation of poroelasticity is not capable of handling situations where porosity is zero in some regions while a significant amount of fluid exists in other parts of the poroelastic body. Specifically, the traditional formulation fails to solve the problem when the porosity becomes sufficiently small, specifically below $\np{5e-4}$ in this case, as shown in Fig.~\ref{fig:ExampleProblem Analysis}\subref{fig:ExampleProblem Traditional Formulation}.
Therefore, in the context of modeling \ac{CSF} circulation and exchange in the \ac{CNS}, these advancements can potentially facilitate the development of computational models capable of handling extreme conditions, such as the spinal cord squeezing caused by abdominal muscle contraction in mice~\cite{Garborg2025Gut-}.

\section{Peristalsis in perivascular space}\label{sec: PVS}

\subsection{Motivation and governing equations}
As mentioned in the introduction, the field of neuroscience focused on the identification of drivers and pathways for the clearance of toxic metabolic by-products has seen significant applications of physics-based fluid transport models to brain research  \cite{Hladky2018Elimination-of-,Hladky2022The-glymphatic-}. Here we revisit a problem of some importance concerning a specific hypothesis that has been strongly debated in the area of waste clearance from the brain. 

Cerebral blood vessels have several features that make them different from blood vessels in other parts of the body (cf., e.g., \cite{Abbott2018The-Role-of-Brain-0,Ballabh2004The-Blood--Brai-0}). One of these features is that said vessels appear to be surrounded by a barrier, almost a sheath, called the glia limitans, consisting of astrocytic end-feet (cf., e.g., \cite{Hladky2022The-glymphatic-,Abbott2012Overview-and-Introduction-0,Iliff2012A-Paravascular-0}). The space between the outer wall of a cerebral blood vessel and the glia limitans is called the \acf{PVS} and it is fluid-rich \cite{Lam2017TheUltrastructu}. Based on a large set of experiments, Nedergaard and co-workers \cite{Iliff2013Brain-wide-Path0,Iliff2013Is-There-a-Cere-0,Iliff2012A-Paravascular-0,Iliff2013Cerebral-Arteri-0} hypothesized that \ac{CSF} is drawn from the subarachnoid space into the \ac{PVS} of surface arterial vessels penetrating into the brain. Furthermore, they hypothesized that wall movements due to heart-gated arterial pulsations could drive significant \ac{CSF} flow (cf. for a careful discussion of what represents significant flow, see \cite{Kedarasetti2020Arterial-Pulsat,Kedarasetti2022ArterialVasodil,Kedarasetti2020Functional-Hype}) in the \ac{PVS} by peristalsis \cite{Iliff2013Cerebral-Arteri-0}.

\Ac{WO} \cite{Wang2011Fluid-Mechanics-0} had studied peristaltic flow in the \ac{PVS} in a very different context, namely that of convection-enhanced drug delivery. From a mathematical viewpoint, after modeling the \ac{PVS} as a porous medium, \ac{WO} derived an elegant analytical solution to a problem involving steady Darcy flow (cf.\ \cite{Coussy2010Mechanics-and-P0}) through the \ac{PVS}. The steady flow field was that perceived by an observer traveling at a constant velocity with the peristaltic wave. The steady flow assumption, paired with the assumptions that the porosity was constant and uniform in the moving observer's frame, allowed \ac{WO} to determine an analytical solution without having to determine the deformation of the underlying solid skeleton.

We became interested in understanding to what extent the work by \ac{WO} could lend support to the hypothesis proposed by Nedergaard and co-workers. Furthermore, since \ac{WO}'s analysis bypassed the determination of the deformation of the solid phase, we became interested in solving the problem numerically, this time assuming that the porosity was uniform in the reference configuration of the solid skeleton, thereby allowing the porosity to change spatially due to deformation. 
We now revisit the problem studied by \ac{WO}, utilizing the poroelastic formulation proposed in this paper, as the commonly used formulation fails to capture the case of zero porosity.

\begin{figure}[ht]
    \centering
    \includegraphics[width=0.6\linewidth]{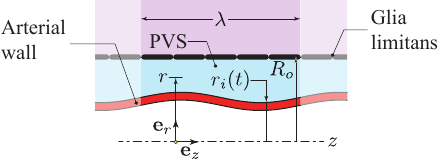}
    \caption{Schematic of the 2D axisymmetric geometry of the revisited WO's problem. The red region depicts the arterial wall undergoing a peristaltic motion with a wavelength $\lambda$ and velocity $c$, propagating in the positive $z$-direction. The blue region illustrates the fluid within the \ac{PVS}. The violet area corresponds to the brain tissue, which is separated from the \ac{PVS} by the glia limitans membrane.}
    \label{fig:PVS}
\end{figure}

Following \ac{WO}, Fig.~\ref{fig:PVS} shows a schematic of the \ac{PVS} surrounding a cerebral artery, with the arterial wall shown in red. The $z$-axis corresponds to the axis of the artery, and axial symmetry around this axis is assumed. The \ac{PVS} is depicted in blue.
Its outer boundary, with radius $R_o$, corresponds to the glia limitans \cite{Abbott2018The-Role-of-Brain-0}. As done by \ac{WO}, we assume that $R_{o}$ is a constant, that is, we treat brain parenchyma as rigid, thus causing the glia limitans to remain stationary.
The inner boundary of the \ac{PVS} coincides with the outer surface of the arterial wall, with radius $r_i(t)$, and its motion is characterized as
\begin{equation}
r_i(t)=h+b \sin \left[\frac{2 \pi}{\lambda}(z-c t)\right],
\label{equ:arterial_wall_disp}
\end{equation}
where $b$ is the wall displacement amplitude, $h$ is the mean (outer) arterial radius, $c$ is the peristaltic wave speed, and $\lambda$ is the wavelength \cite{Wang2011Fluid-Mechanics-0}.
To account for both the deformation of the fluid pathway and the fluid flow within the \ac{PVS}, we adopt the modified version of the poroelastic model described in Section \ref{sec: reformulation}. Since the peristaltic wave moves at a constant velocity, an observer traveling with the wave can be treated as inertial. Therefore, we reformulate the steady-state poroelastic model by representing it in a moving reference frame.

\begin{figure}[hbt]
    \begin{subfigure}[b]{0.3\textwidth}
	\centering
	\includegraphics[height=1.1in]{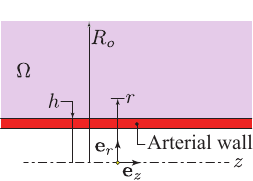}
	\caption{}
	\label{fig:PVSSchematicRest}
    \end{subfigure}
    \begin{subfigure}[b]{0.3\textwidth}
	\centering
	\includegraphics[height=1.1in]{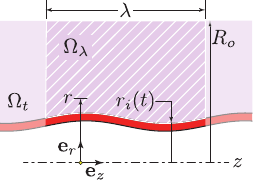}
	\caption{}
	\label{fig:PVSSchematicDeformed}
    \end{subfigure}
    \begin{subfigure}[b]{0.38\textwidth}
	\centering
	\includegraphics[height=1.1in]{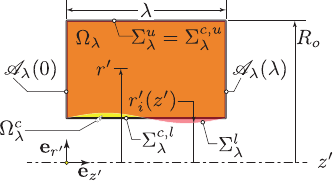}
	\caption{}
	\label{fig:PVSSchematicComputational}
    \end{subfigure}
    \caption{\subref{fig:PVSSchematicRest} The rest state of the poroelastic domain $\Omega$, which can be either \ac{PVS} only or union of \ac{PVS} and brain parenchyma. \subref{fig:PVSSchematicDeformed} The deformed poroelastic domain $\Omega_t$. The region $\Omega_{\lambda}\subset \Omega_t$ is a finite control volume that moves alongside the observer, tracking the peristaltic wave, which is assumed to propagate in the positive $z$-axis direction at velocity $c$. \subref{fig:PVSSchematicComputational} The deformed domain in orange and the computational domains in yellow seen by the moving reference frame.}
    \label{fig:PVS_Configs}
\end{figure}

As illustrated in Fig.~\ref{fig:PVS_Configs}\subref{fig:PVSSchematicRest}, we examine the deformation of the tissue around a blood vessel. Initially, the material fills a region $\Omega$ with a nominal internal radius $h$ and an outer radius $R_o$. To simplify the model, the entire region $\Omega$ is treated as a homogeneous poroelastic medium. Depending on the specific value of $R_o$ and the variation in porosity across the region, the domain $\Omega$ can be interpreted in different ways: as the \ac{PVS}, as the tissue around the vessel without a clearly defined \ac{PVS}, or as the combined \ac{PVS} and surrounding tissue.
The region $\Omega_t$ (Fig.~\ref{fig:PVS_Configs}\subref{fig:PVSSchematicDeformed}) in the cylindrical coordinate system $(r, \theta, z)$ represents the deformed poroelastic domain after the peristaltic wave has propagated in the positive $z$-direction. A point $\bv{x} = r {\bv{e}}_r(\theta) + z {\bv{e}}_z$ in $\Omega_t$ satisfies the conditions $r_i(t) < r < R_o$ and $-\infty < z < \infty$.
As depicted in Fig.~\ref{fig:PVS_Configs}\subref{fig:PVSSchematicComputational}, we reformulate the problem in a cylindrical coordinate system within the moving frame, defined by coordinates $(r^\prime, \theta^\prime, z^\prime)$. The $z^\prime$ axis aligns with the $z$ axis, pointing in the same direction, and these coordinates are connected to those of the stationary frame through the following transformation:
\begin{equation}
\begin{aligned}
r^{\prime}=r, \quad \theta^{\prime}=\theta, \quad \text { and } \quad z^{\prime}=z-c t .
\end{aligned}
\end{equation}
A time-independent control volume $\Omega_{\lambda}$ for the moving observer can be defined as
\begin{equation}
\begin{aligned}
\Omega_\lambda \coloneqq \left\{\bv{x}^{\prime} \mid r^{\prime}\left(\bv{x}^{\prime}\right) \in\bigl[r_i^{\prime}\left(z^{\prime}\right), R_o\bigr], \theta^{\prime} \in[0,2 \pi], z^{\prime} \in[0, \lambda]\right\},
\end{aligned}
\end{equation}
where $r_i^\prime(z^\prime)$ is obtained from Eq.~\eqref{equ:arterial_wall_disp} by setting $z-c t=z^\prime$.

We define the boundary segments of $\Omega_\lambda$ at $z^\prime = \np{0}$ and $z^\prime = \lambda$ as $\mathscr{A}_\lambda(0)$ and $\mathscr{A}_\lambda(\lambda)$, respectively. It is important to note that these two boundary segments, $\mathscr{A}_\lambda(0)$ and $\mathscr{A}_\lambda(\lambda)$, are homologous. This means that one can be perfectly aligned with the other via a simple translation along the $z^\prime$-axis by an amount of $\lambda$. A function $\phi(r^\prime, \theta^\prime, z^\prime)$ defined over $\Omega_\lambda$ is said to be $\lambda$-periodic if it satisfies
$
\phi(r^\prime, \theta^\prime, 0) = \phi(r^\prime, \theta^\prime, \lambda)
$
for all values of $r^\prime \in [r^\prime_{i}(0), R_o]$ and $\theta^\prime \in [0, 2\pi]$. For convenience, we denote by $\Sigma_\lambda^l$ and $\Sigma_\lambda^u$ the lower and upper portions of the boundary of $\Omega_\lambda$ at $r^\prime = r^\prime_i(z)$ and $r^\prime = R_o$, respectively.

In the moving reference frame, physical quantities will be labeled with a prime to differentiate them from those in the stationary reference frame.
We suggest that the moving observer experiences the fields $\bv{v}_{\s}^{\prime}, \bv{v}^{\prime}_{\f}, \Grad{p}^{\prime}$, and $\ts{F}^{\prime}_{\s}$ (and thus $J_{\s}^{\prime}$) as being both time-independent and periodic with a period of $\lambda$.
Based on these conditions, the displacement field $\bv{u}_\s^\prime$ can be expressed as $\bar{\bv{u}}_\s^\prime-c t \bv{e}_{z^\prime}$, where $\bar{\bv{u}}_\s^\prime$ is a $\lambda$-periodic function mapping from $\Omega_\lambda$ to the space $\mathscr{V}$.
Moreover, following \ac{WO}, we define $p^\prime = \bar{p}^\prime + \left( \Delta p/\lambda \right) z^\prime \bv{e}_{z^\prime}$, where $\bar{p}^\prime: \Omega_\lambda \to \mathbb{R}$ is a function that is periodic with period $\lambda$, and $\Delta p \in \mathbb{R}$ is a \emph{given} constant.
Accounting for the relationship between the stationary and moving frames, we have that the velocity field of the solid phase as observed by the moving frame is given by 
\begin{equation}
\begin{aligned}
\bv{v}^\prime_\s : \Omega_\lambda \to \mathscr{V}, \quad \bv{v}^\prime_\s = -c\,\ts{F}^\prime_\s \, \bv{e}_{z^\prime}.
\end{aligned}
\end{equation}
Since the moving frame is inertial, the governing equations in this frame are mathematically equivalent to those in Eqs.~\eqref{equ:mass_balance_mixture}, \eqref{equ:eom_mixture}, and~\eqref{equ:eom_fluid_mod}, with the only difference being that the unprimed quantities are replaced by their corresponding primed counterparts.
Choosing $\bar{\bv{u}}_\s^\prime$, $\bv{v}_\s^\prime$, $\bv{v}_\f^\prime$, and $\bar{p}^\prime$  as the primary unknowns, the following boundary conditions are imposed
\begin{alignat}{2}
\bar{\bv{u}}_{\s}^{\prime} & =b \sin \biggl(\frac{2 \pi z^{\prime}}{\lambda}\biggr)\, \bv{e}_{r^{\prime}} &\quad&\text {on $\Sigma_\lambda^l$,}
\\
\bar{\bv{u}}_{\s}^{\prime} &= \bv{0} &\quad&\text {on $\Sigma_\lambda^u$,}
\\
(\bv{v}_{\f}^{\prime}-\bv{v}_{\s}^{\prime}) \cdot \bv{n}^{\prime} &= 0 &\quad&\text{on $\Sigma_\lambda^l \cup \Sigma_\lambda^u$,}
\end{alignat}
where $\bv{n}^\prime$ represents the outward unit normal to the boundary of $\Omega_\lambda$. 
As stated, the problem has a unique solution for $\bar{p}^\prime$ up to an arbitrary constant. We therefore impose an additional scalar constraint requiring that the average value of $\bar{p}^\prime$ be zero.

Although the problem can be solved directly on the domain $\Omega_\lambda$, we opt to use the same \ac{ALE} approach discussed earlier in the paper. This involves transforming the problem onto a domain with a simpler geometry, specifically a right hollow cylinder.
To achieve this, we define a control volume $\Omega_\lambda^c$, which consists of points $\hat{\bv{x}}^\prime$ that satisfy the following conditions
\begin{equation}
\Omega_\lambda^c \coloneqq\left\{\hat{\bv{x}}^{\prime} \mid r^{\prime}\left(\hat{\bv{x}}^{\prime}\right) \in\left[h, R_o\right], \theta^{\prime}\left(\hat{\bv{x}}^{\prime}\right) \in[0,2 \pi], z^{\prime}\left(\hat{\bv{x}}^{\prime}\right) \in[0, \lambda]\right\}.
\end{equation}
We can reinterpret the field $\bar{\bv{u}}_{\s}^{\prime}$ as the displacement field that transforms the region $\Omega_\lambda$ into $\Omega_\lambda^c$. Specifically, consider the diffeomorphism $\bv{\zeta}(\bv{x}^\prime) : \Omega_\lambda \to \Omega_\lambda^c$, which is defined by
\begin{equation}
\hat{\bv{x}}^{\prime} = \bv{\zeta}(\bv{x}^{\prime}) \coloneqq \bv{x}^{\prime} - \bar{\bv{u}}_{\s}^{\prime}.
\end{equation}
Its inverse describes the deformation that transforms the domain $\Omega_\lambda^c$ into $\Omega_\lambda$. We refer to this inverse map as $\bv{\chi}_\lambda$, and we use it to reframe our problem in terms of the computational domain $\Omega_\lambda^c$. In this context, we can express the map $\bv{\chi}_\lambda$ as follows
\begin{equation}
\bv{\chi}_\lambda: \Omega_\lambda^c \to \Omega_\lambda, \quad \bv{\chi}_\lambda\left(\hat{\bv{x}}^{\prime}\right) \coloneqq \bv{\zeta}^{-1}\left(\hat{\bv{x}}^{\prime}\right).
\end{equation}
Next, we define the deformation gradient $\ts{F}_\lambda$ and the Jacobian determinant $J_\lambda$ as
\begin{equation}
\ts{F}_\lambda
\coloneqq
\Grad{\bv{\chi}}_\lambda
\quad\text{and}\quad
J_\lambda
\coloneqq \operatorname{det} \ts{F}_\lambda.
\end{equation}
These quantities are crucial for describing the mapping and the corresponding changes in volume and geometry as we move from $\Omega_\lambda^c$ to $\Omega_\lambda$. Therefore, letting $\circd{s}$ and $\circd{s}$ denote $\Omega_\lambda^c$ to $\Omega_\lambda$, respectively, we have
\begin{equation}
\hat{\bv{u}}^\prime
=
\bigl(\bar{\bv{u}}_\s^\prime\bigr)^{\circd{s}}
=\bv{\chi}_\lambda
-\hat{\bv{x}}^\prime,
\quad
(\ts{F}^{\prime}_\s)^{\circd{s}}
=\ts{F}_\lambda
,
\quad\text{and}\quad
(J^{\prime}_\s)^{\circd{s}}
= J_\lambda
.
\end{equation}
where $\hat{\bv{u}}^\prime$ is the Lagrangian representation of the solid's displacement field. The Lagrangian form of the solid velocity is also defined as
\begin{equation}
\hat{\bv{v}}^\prime_\s
=
(\bv{v}^\prime_\s)^{\circd{s}}
\end{equation}

As already mentioned in the previous section, we present equations, with the expectation that we will take advantage of the \ac{ALE} support provided by \ac{CMPH}~\cite{COMSOLCite}. This said, our problem  consists in finding
$\lambda$-periodic fields $\hat{\bv{u}}_\s^\prime\left(\hat{\bv{x}}^\prime\right)$, $\hat{\bv{v}}_\s^\prime\left(\hat{\bv{x}}^\prime\right)$, $\bv{v}_\f^\prime\left(\bv{x}^\prime\right)$, and $\bar{p}^\prime\left(\bv{x}^{\prime}\right)$, such that
\begin{align}
\label{eq: vel consistency}
\hat{\bv{v}}_{\s}^{\prime} + c \,\ts{F}_\lambda \, \bv{e}_{\hat{z}^{\prime}}&=\bv{0} \quad\text{in $\Omega_\lambda^c$},
\\
\begin{multlined}[b]
\rho_{\f}^*(J_\lambda-\phi_{R_\s})(-\bv{b}_{\f}+\Grad{\bv{v}_{\f}^{\prime}}[\bv{v}_{\f}^{\prime}])^{\circd{s}}
\\
-\rho_{\s}^* \phi_{R_\s}\left(\bv{b}_{\s}^{\circd{s}}+c \, \Grad{\hat{\bv{v}}_{\s}^{\prime}} \left[\bv{e}_{\hat{z}^{\prime}}\right]\right)+
J_\lambda
(\Grad{p^\prime})^{\circd{s}}
-\Div{\ts{P}^e}
\end{multlined}
& =\bv{0} \quad\text{in $\Omega_\lambda^c$},
\\
\frac{\phi_\f \rho_{\f}^*}{\eta(\phi_\f)}\left(-\bv{b}_{\f}+
\Grad{\bv{v}_{\f}^{\prime}}\left[\bv{v}_{\f}^{\prime}\right]\right)+\frac{\phi_\f}{\eta(\phi_\f)} \Grad{p^{\prime}}+\left(\bv{v}_{\f}^{\prime}-(\hat{\bv{v}}_{\s}^{\prime})^{\circd{e}}\right) &=\bv{0} \quad\text{in $\Omega_\lambda$},
\\
\Div{\left(\phi_\s (\hat{\bv{v}}_\s^\prime)^{\circd{e}}+\phi_{\f} \bv{v}_{\f}^{\prime}\right)} &=\bv{0} \quad\text{in $\Omega_\lambda$},
\end{align}
and such that
\begin{alignat}{2}
\label{eq: disp. BC wave}
\hat{\bv{u}}_{\s}^{\prime} &= b \sin \biggl(\frac{2 \pi \hat{z}^{\prime}}{\lambda}\biggr) \, \bv{e}_{r^{\prime}}
&\quad&\text{on $\Sigma_\lambda^{c,l}$,}
\\
\label{eq: disp. BC upper}
\hat{\bv{u}}_{\s}^{\prime} &= \bv{0}
&\quad&\text{on $\Sigma_\lambda^{c,u}$,}
\\
\label{eq: vel. BC}
\left(\bv{v}_{\f}^{\prime}-(\hat{\bv{v}}_{\s}^{\prime})^{\circd{e}}
\right) \cdot \bv{n}^{\prime} &= 0 &\quad&\text{on $\Sigma_\lambda^{l} \cup \Sigma_\lambda^{u}$.}
\end{alignat}
The weak form of the above equations is the same as that used in the previous section.

\subsection{Results}\label{sec: results}
We solve the problem in Eqs.~\eqref{eq: vel consistency}--\eqref{eq: vel. BC}, using the parameter values indicated in Table~\ref{tab:WO params}.
\begin{table}[hbt]
\centering
\caption{Parameters for peristaltic wave modeling as in  \cite{Wang2011Fluid-Mechanics-0}, along with values for the other parameters needed by our formulation.}
\label{tab:WO params}
\begin{threeparttable}[b]
\begin{tabular}{lll}
	\hline
        Quantity & Description & Value \\
	\hline
        $c$ & Wave speed & $\np[m/s]{1}$ \\
	$h$ & Base inner radius of \ac{PVS} annulus & $\np[\mu m]{10}$ \\
	$R_o$ & Outer radius of annulus & $\np[\mu m]{11}$ \\
	$\lambda$ & Wavelength of peristaltic wave & $\np[m]{0.2}$\tnote{$\dagger$} \\
	$\phi_{R_\f}$ & Referential porosity & 0.2 \\
	$\mu_\f$ & \ac{CSF} viscosity & $\np[Pa\cdot s]{9e{-4}}$ \\
	$\overline{\kappa}_\s$ & Permeability of tissue &  $\np[m^2]{1.8e{-14}}$ \\
	$b$ & Amplitude of peristaltic wave & $\np[\mu m]{0.25}$ \\
        $\rho^*_\s$ & Solid true density\tnote{$\ddagger$} & $\np[kg/m^3]{1e3}$ \cite{Kedarasetti2022ArterialVasodil}\\
        $\rho^*_\f$ & Fluid true density\tnote{$\ddagger$} & $\np[kg/m^3]{1e3}$ \cite{Kedarasetti2022ArterialVasodil}\\
        $\phi_{\f_\mathrm{c}}$ & Critical porosity\tnote{$\ddagger$} & $\np{0.1}$\\
        $\mu_\s$ & Solid shear modulus\tnote{$\ddagger$} & $\np[kPa]{14}$ \cite{Miller2011BiomechanicsBrain0}\\
	\hline
\end{tabular}
\begin{tablenotes}
       \item [$\dagger$] {\footnotesize This value corresponds to a heart rate $f$ of $\np[Hz]{5}$.}
       \item [$\ddagger$] {\footnotesize This quantity is not reported nor used by \cite{Wang2011Fluid-Mechanics-0}, and is estimated or taken from other sources.}
\end{tablenotes}
\end{threeparttable}
\end{table}
These parameter values include those indicated in \ac{WO}. Clearly, as the solution in \ac{WO} is only based on the Darcy flow model, as opposed to our poroelastic setting, Table~\ref{tab:WO params} includes parameters needed by our model that do not appear in \ac{WO}.

The computational domain is discretized using a uniform mesh consisting of identical rectangular elements. For all results, a characteristic mesh size of $h_e=\np[mm]{0.1}$ is used, with a total of $\np{50}$ elements distributed along the radial direction.
Except for the nonlinear method, which was chosen to be a Newton's method with a constant damping factor of $1$ and minimal Jacobian updates, all other settings, such as the element type and solver configurations, follow those specified in Section \ref{subsec:mms} for the method of manufactured solutions.

Here, we address three distinct subproblems. First, we investigate the \ac{PVS} alone, assuming a rigid brain---similar to the approach taken in \cite{Wang2011Fluid-Mechanics-0}---and examine the effect of domain length on the pressure and volume flow rate fields. The second subproblem extends the domain (by considering a larger value of $R_o$) to include both the \ac{PVS} and the brain parenchyma, under the assumption of uniform referential porosity. Finally, in the third subproblem, we explore a scenario where the referential description of the porosity is heterogeneous across the domain, and accomplish the same set of analyses. 

\begin{figure}[hbt]
    \begin{subfigure}[b]{0.5\textwidth}
        \centering
        \includegraphics[width=\textwidth]{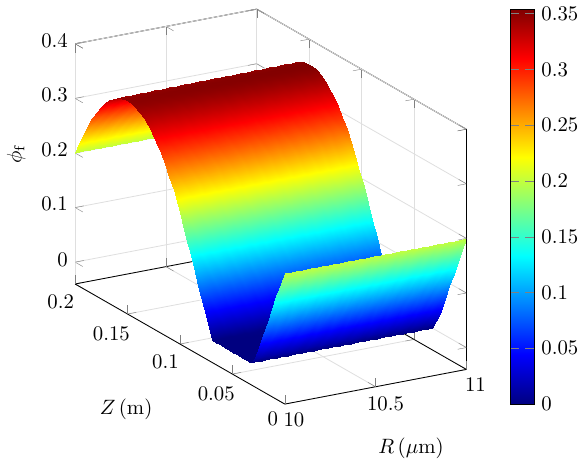}
        \caption{}
        \label{fig:res_zero_porosity_mus14kPa}
    \end{subfigure}
    \begin{subfigure}[b]{0.5\textwidth}
        \centering
        \includegraphics[width=\textwidth]{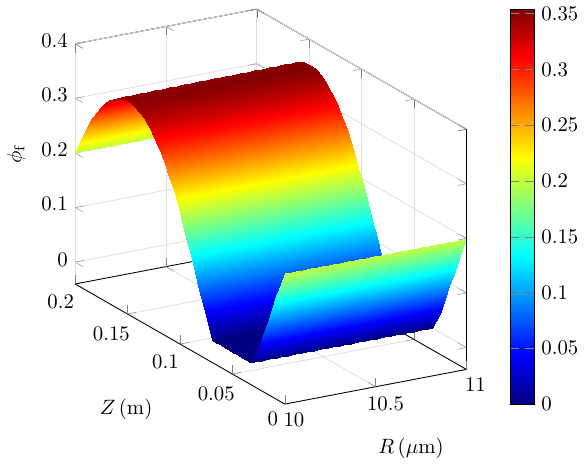}
        \caption{}
        \label{fig:res_zero_porosity_mus2kPa}
    \end{subfigure}
    \caption{Porosity $\phi_{\f}$ (shown in the reference configuration of the solid component) approaches zero in some regions when the peristaltic wave is applied, indicating that the redistribution of the fluid component is highly affected by the deformation of the solid component. Obtained using the parameters from \subref{fig:res_zero_porosity_mus14kPa} subproblem 1 and \subref{fig:res_zero_porosity_mus2kPa} subproblem 1 with $\mu_\s=\np[kPa]{2}$ where all other parameters are identical. For this problem, the porosity becomes zero regardless of the shear modulus value. The scaling of radial and axial axes is adjusted differently for visualization purposes.}
	\label{fig:res_zero_porosity}
\end{figure}

The first result we present is the porosity distribution in the solution domain, which, based on the chosen value of $R_o$ (Table~\ref{tab:WO params}), represents the \ac{PVS} alone (see Fig.~\ref{fig:res_zero_porosity}).
With the geometry and peristaltic wave amplitude indicated in \ac{WO}, the \ac{PVS} is characterized by a region of zero porosity, that is a region from which the fluid phase has been completely squeezed out, as shown in Fig.~\ref{fig:res_zero_porosity}. This result cannot be obtained within \ac{WO}'s framework because the latter does not model the deformation of the solid phase and the effect of this deformation on the porosity of the system. This effect can only be captured by a poroelastic model that accounts for large deformations. Furthermore, we wish to mention that we attempted to solve this same physical problem with the formulation in Section~\ref{sec: traditional formulation}, i.e., a formulation not specifically designed to remain valid when the porosity vanishes. The attempts in questions resulted in the same type of failure described in relation to Fig.~\ref{fig:ExampleProblem Analysis}. We could obtain sequences of solutions for increasing values of the peristaltic wave amplitude, but we could never achieve the value of $\sfrac{1}{4}$ of \ac{PVS} lumen width indicated in \ac{WO} \cite{Wang2011Fluid-Mechanics-0}. The failure was preceded by the achievement of negative values of porosity.

What we can immediately conclude is that, to begin to have physiological relevance, a flow model in the \ac{PVS} needs to account for the deformation of the \ac{PVS}. It also needs to account for the deformation of the brain tissue induced by a the motion of the glia limitans. We also conclude that, while we do not expect physiological situations in which all fluid is expelled from tissue, a modeling framework able to numerically handle these cases represents a useful investigative tool in that it allows us to more easily identify modeling assumptions that lead to extreme conditions.

\begin{figure}[p]
    \centering
    \begin{subfigure}{0.3\linewidth}{\includegraphics[width=\textwidth]{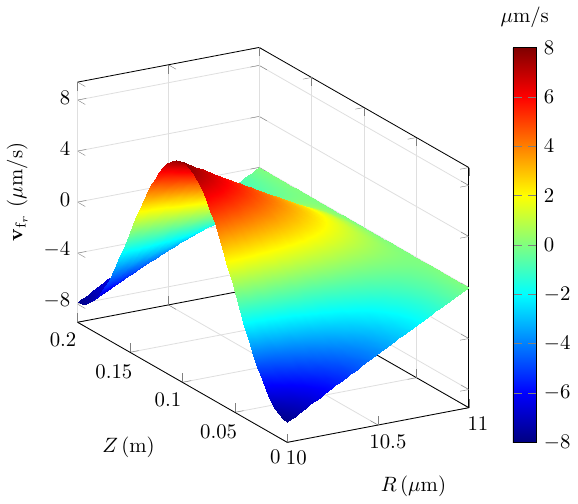}\caption{}\label{fig:ResPVSVfrC1Ro11}}\end{subfigure}\hfill
    \begin{subfigure}{0.3\linewidth}{\includegraphics[width=\textwidth]{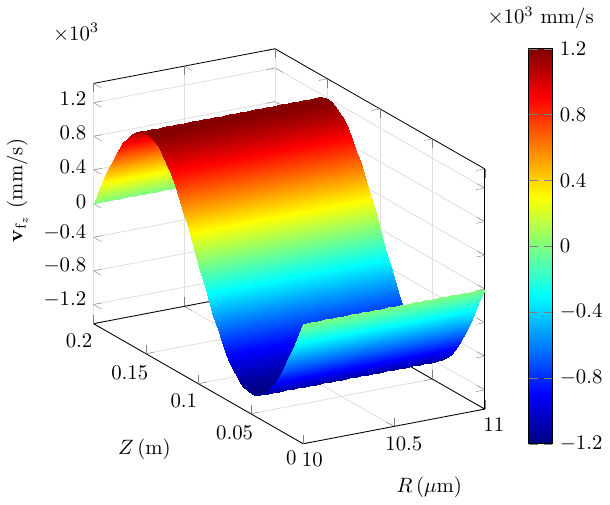}\caption{}\label{fig:ResPVSVfzC1Ro11}}\end{subfigure}\hfill
    \begin{subfigure}{0.3\linewidth}{\includegraphics[width=\textwidth]{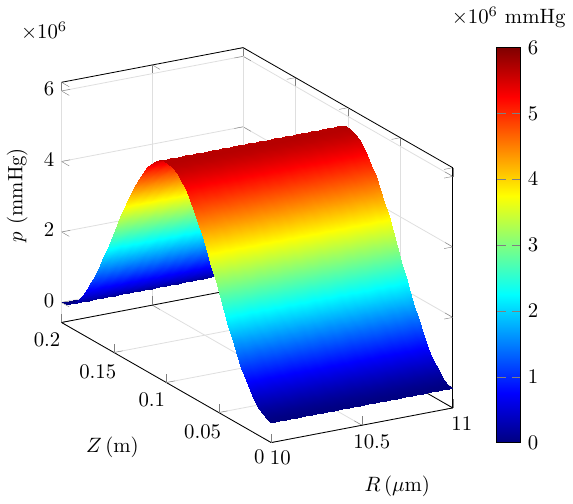}\caption{}\label{fig:ResPVSPC1Ro11}}\end{subfigure}
    
    \begin{subfigure}{0.3\linewidth}{\includegraphics[width=\textwidth]{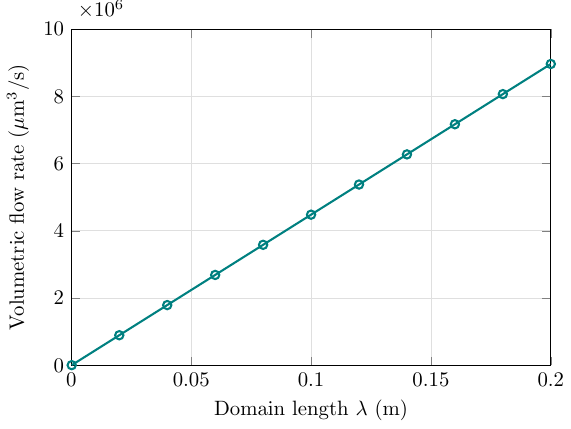}\caption{}\label{fig:ResPVSVFRRo11}}\end{subfigure}\hfill
    \begin{subfigure}{0.3\linewidth}{\includegraphics[width=\textwidth]{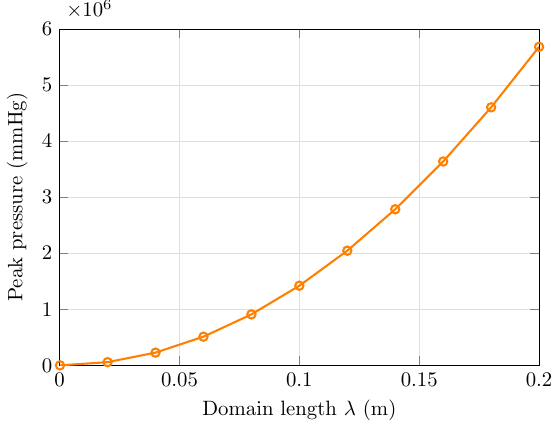}\caption{}\label{fig:ResPVSPPRo11}}\end{subfigure}\hfill
    \begin{subfigure}{0.3\linewidth}{\includegraphics[width=\textwidth]{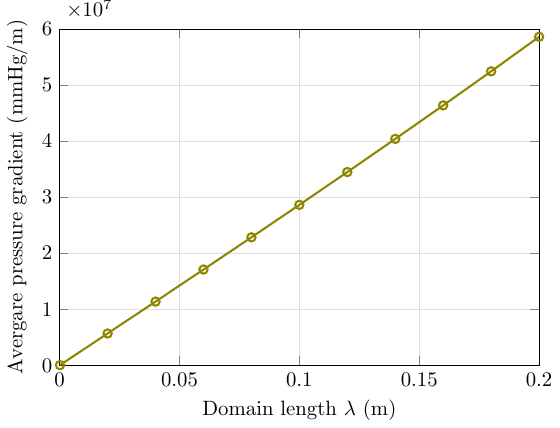}\caption{}\label{fig:ResPVSAPGRo11}}\end{subfigure}
    
    \begin{subfigure}{0.3\linewidth}{\includegraphics[width=\textwidth]{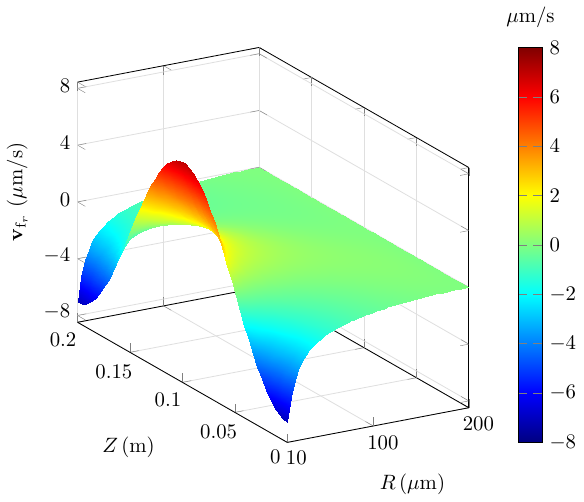}\caption{}\label{fig:ResPVSVfrC1Ro200}}\end{subfigure}\hfill
    \begin{subfigure}{0.3\linewidth}{\includegraphics[width=\textwidth]{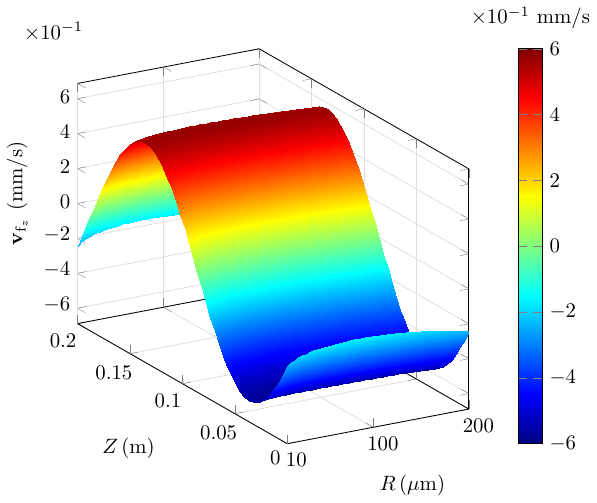}\caption{}\label{fig:ResPVSVfzC1Ro200}}\end{subfigure}\hfill
    \begin{subfigure}{0.3\linewidth}{\includegraphics[width=\textwidth]{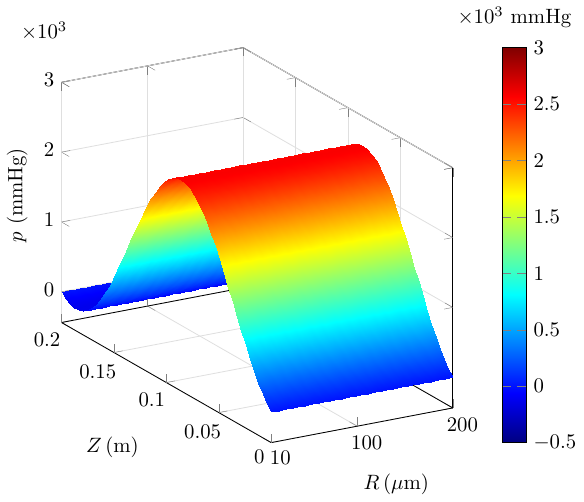}\caption{}\label{fig:ResPVSPC1Ro200}}\end{subfigure}
    
    \begin{subfigure}{0.3\linewidth}{\includegraphics[width=\textwidth]{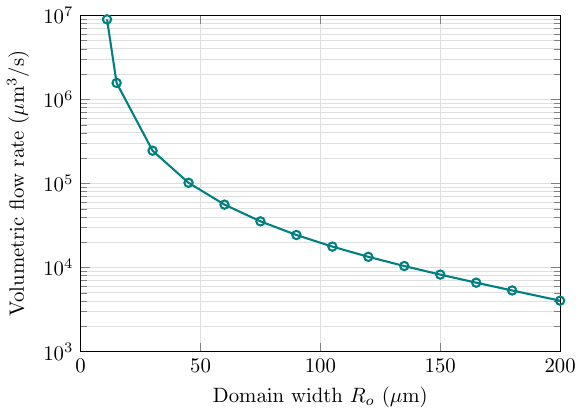}\caption{}\label{fig:ResPVSVFRC1}}\end{subfigure}\hfill
    \begin{subfigure}{0.3\linewidth}{\includegraphics[width=\textwidth]{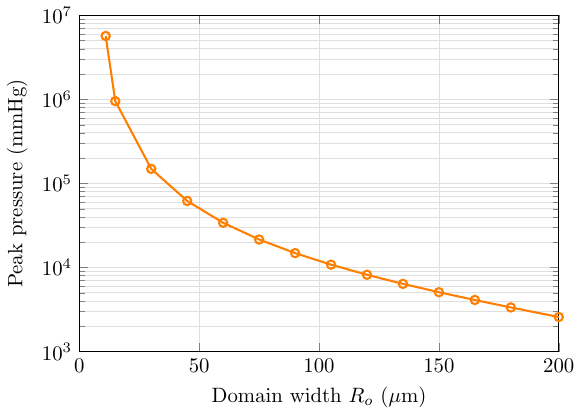}\caption{}\label{fig:ResPVSPPC1}}\end{subfigure}\hfill
    \begin{subfigure}{0.3\linewidth}{\includegraphics[width=\textwidth]{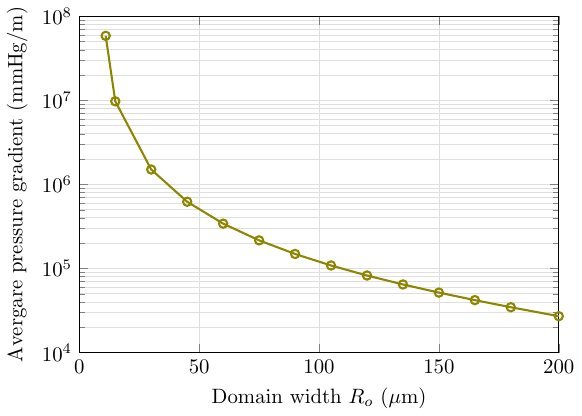}\caption{}\label{fig:ResPVSAPGC1}}\end{subfigure}

    \caption{Subproblem 1: numerical solution of the fluid mechanics problem in the \ac{PVS} when the brain is assumed to be rigid, as outlined by WO. \subref{fig:ResPVSVfrC1Ro11} Radial fluid velocity, \subref{fig:ResPVSVfzC1Ro11} axial fluid velocity, and \subref{fig:ResPVSPC1Ro11} pressure distribution in the stationary frame, for the parameters specified in Table~\ref{tab:WO params}. \subref{fig:ResPVSVFRRo11} Volume flow rate, \subref{fig:ResPVSPPRo11} peak pressure, and \subref{fig:ResPVSAPGRo11} average pressure gradient as a function of domain length  $\lambda$. 
    Subproblem 2: numerical solution of the fluid mechanics problem in a seamless \ac{PVS}--brain domain with uniform referential porosity.
    \subref{fig:ResPVSVfrC1Ro200} Radial fluid velocity, \subref{fig:ResPVSVfzC1Ro200} axial fluid velocity, and \subref{fig:ResPVSPC1Ro200} pressure distribution in the stationary frame, for the parameters listed in Table \ref{tab:WO params}, with the exception of $R_o = \np[\mu m]{200}$. \subref{fig:ResPVSVFRC1} Volume flow rate, \subref{fig:ResPVSPPC1} peak pressure, and \subref{fig:ResPVSAPGC1} average pressure gradient as a function of the outer radius $R_o$. Scaling of radial and axial axes differs for visualization purposes.}
    \label{fig:ResPVS}
\end{figure}

The distribution of the fluid velocity components and pressure for the first subproblem (rigid brain case), obtained by solving the refined poroelastic model using the parameters in Table~\ref{tab:WO params}, is presented in Figs.~\ref{fig:ResPVS}\subref{fig:ResPVSVfrC1Ro11}--\subref{fig:ResPVSPC1Ro11}, respectively. The peak pressure and the average pressure gradient are $\np[mmHg]{5.68e6}$ and $\np[mmHg/m]{5.86e7}$, respectively, while the volume flow rate, defined as
\begin{equation}
    \bar{Q} = \dfrac{1}{\lambda}\int_{\Omega_\lambda^c} \bv{v}_{\flt_z} J_\lambda\ d\Omega_\lambda^c,
\end{equation}
is $\np[\mu m^3/s]{8.96e6}$. The simulations were conducted for a range of domain lengths, with the wave speed adjusted accordingly to maintain consistency with the formulation and, specifically, the periodicity assumption. From Figs.~\ref{fig:ResPVS}\subref{fig:ResPVSVFRRo11}--\subref{fig:ResPVSAPGRo11}, it can be observed that both the volume flow rate and average pressure gradient are proportional to the domain length.
The lowest volume flow rate, peak pressure, and average pressure gradient are achieved with a domain length of $\np[\mu m]{150}$, where the values are $\np[\mu m^3/s]{6.72e3}$, $\np[mmHg]{3.2}$, and $\np[mmHg/m]{4.26e4}$, respectively.

Based on the analysis using the suggested parameters, the attained values, particularly the peak pressure, are sufficiently high compared to normal intracranial pressure, justifying the use of a poroelastic model for the \emph{deformable} brain parenchyma instead of a rigid brain model, as previously employed by WO. Moreover, the substantial pressure gradients observed suggest that fluid may flow beyond the \ac{PVS} into the surrounding tissue. 
However, in the results presented above, our investigations were limited to the \ac{PVS} immediately adjacent to penetrating blood vessels, with the simplifying assumption of a fixed outer boundary.

To evaluate this idea, as the second subproblem, we explore the behavior of our computational model for a range of values of the outer radius $R_o$ from $\np[\mu m]{11}$ to $\np[\mu m]{200}$, the latter approximating the typical distance between blood vessel centers. 
Indeed, in this subproblem, we treat the brain and \ac{PVS} as a seamless domain with uniform referential porosity, and vary the domain width (i.e., $R_o$) to investigate the quantities of interest.
All other simulation parameters are held constant as stated in Table~\ref{tab:WO params}.
Specifically, Figs.~\ref{fig:ResPVS}\subref{fig:ResPVSVfrC1Ro200}--\subref{fig:ResPVSPC1Ro200} detail the velocity and pressure fields for the scenario with a $\np[\mu m]{200}$ outer radius.
Our results demonstrate that both velocity and pressure distributions are sensitive to variations in the domain’s size. Figures~\ref{fig:ResPVS}\subref{fig:ResPVSVFRC1}--\subref{fig:ResPVSAPGC1} reveal that both the average pressure gradients and the volume flow rates decrease sharply as the outer radius increases.
For an outer radius of $\np[\mu m]{200}$, the corresponding volume flow rate, peak pressure, and average pressure gradient are $\np[\mu m^3/s]{4.03e3}$, $\np[mmHg]{2.58e3}$, and $\np[mmHg/m]{2.71e4}$, respectively.
At the specified volume flow rate, a parenchymal region extending $\np[\mu m]{150}$ in length would be cleared within a $\np[hour]{1.3}$ interval.

This last set of results is based on the assumption that \ac{PVS} and brain parenchyma have the same (uniform) referential porosity. To account for the fact that the \ac{PVS} is richer in fluid than brain tissue, we can adopt a spatially variable referential porosity, as considered in the following third subproblem. In this approach, higher porosity values are assigned near the vessel walls, while a lower constant value is used within the brain parenchyma. We then propose a model in which the referential porosity is defined as
\begin{equation}
\phi_{R_\f}= \begin{cases}\frac{1}{2}\left(\phi_{\f_\infty}+\phi_{\f_0}\right)-\frac{1}{2}\left(\phi_{\f_{\infty}}-\phi_{\f_0}\right) \cos \left(\pi \dfrac{R-h}{\bar{R}-h}\right) & h \leq R \leq \bar{R}, \\ \phi_{\f_{\infty}} & \bar{R}< R \leq R_o, \end{cases}
\end{equation}
where
\begin{equation*}
\phi_{\f_\infty}=\np{0.2}, \quad {\phi_{\f_0}}=\np{0.8}, \quad \text { and } \quad \bar{R}=\np[\mu m]{12}.
\end{equation*}
The value of $h$ is the same as that reported in Table~\ref{tab:WO params}, while $R_o$ varies between $\np[\mu m]{15}$ and $\np[\mu m]{200}$. 
A more detailed approach would involve modeling the \ac{PVS} and the brain parenchyma as two distinct domains separated by a well-defined interface, taking into account the glia limitans barrier. However, our objective in this study is to gain insights into the effects of heterogeneous porosity; therefore, this simplified representation of porosity is sufficient to provide the necessary perspective.

\begin{figure}[ht!]
    \centering
    \begin{subfigure}{0.3\linewidth}{\includegraphics[width=\textwidth]{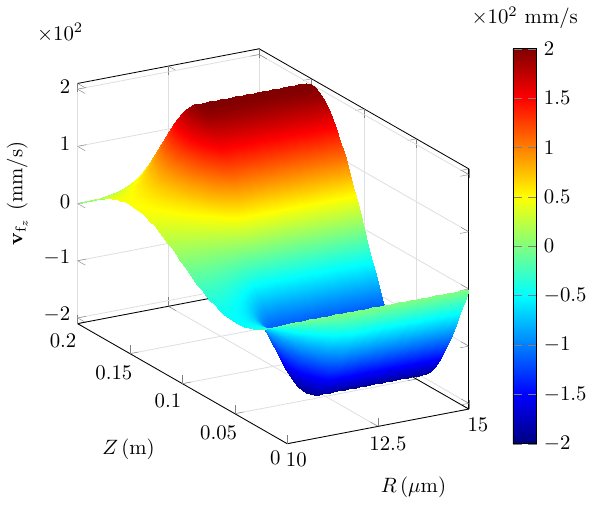}\caption{}\label{fig:ResHetPVSVfzC1Ro15}}\end{subfigure}\hfill
    \begin{subfigure}{0.3\linewidth}{\includegraphics[width=\textwidth]{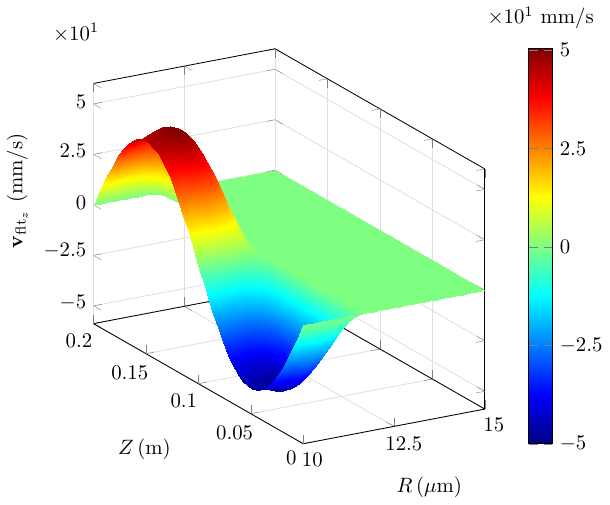}\caption{}\label{fig:ResHetPVSVfltzC1Ro15}}\end{subfigure}\hfill
    \begin{subfigure}{0.3\linewidth}{\includegraphics[width=\textwidth]{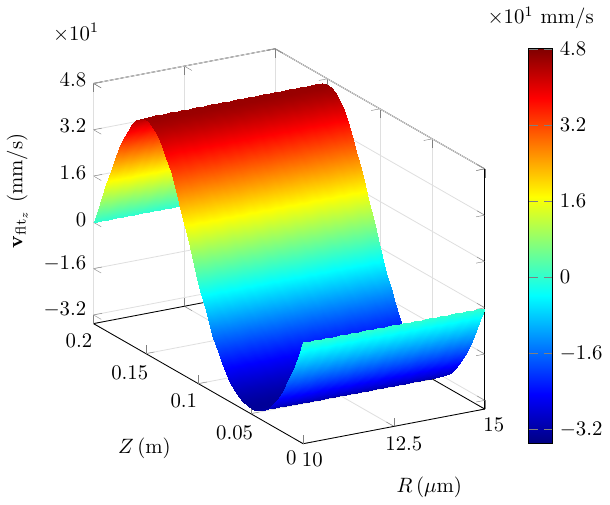}\caption{}\label{fig:ResPVSVfltzC1Ro15}}\end{subfigure}\\[1ex]
    
    \begin{subfigure}{0.3\linewidth}{\includegraphics[width=\textwidth]{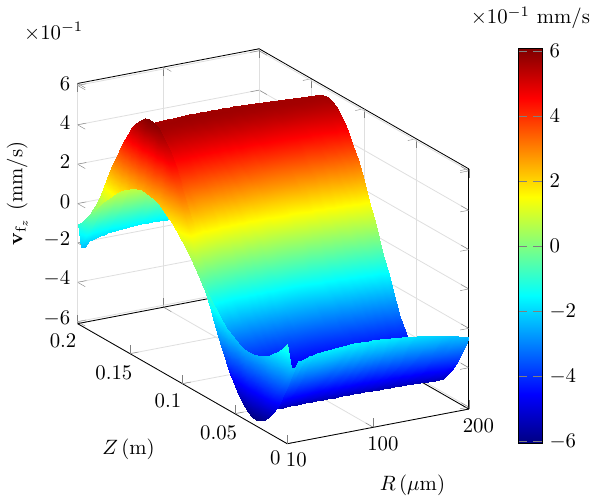}\caption{}\label{fig:ResHetPVSVfzC1Ro200}}\end{subfigure}\hfill
    \begin{subfigure}{0.3\linewidth}{\includegraphics[width=\textwidth]{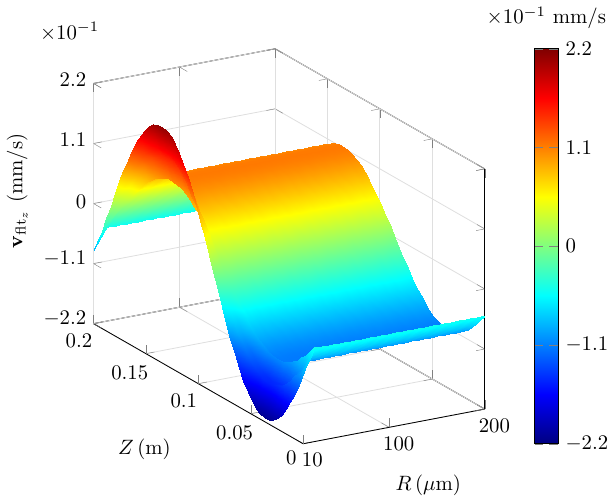}\caption{}\label{fig:ResHetPVSVfltzC1Ro200}}\end{subfigure}\hfill
    \begin{subfigure}{0.3\linewidth}{\includegraphics[width=\textwidth]{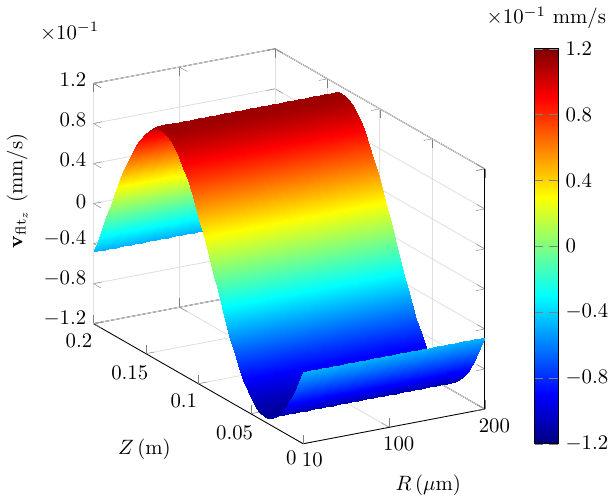}\caption{}\label{fig:ResPVSVfltzC1Ro200}}\end{subfigure}\\[1ex]
        
    \begin{subfigure}{0.3\linewidth}{\includegraphics[width=\textwidth]{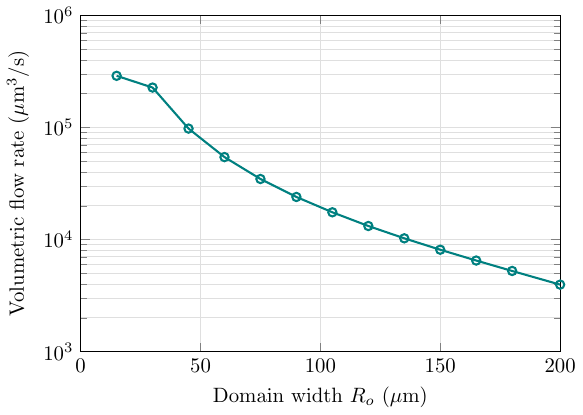}\caption{}\label{fig:ResHetPVSVFRC1}}\end{subfigure}\hfill
    \begin{subfigure}{0.3\linewidth}{\includegraphics[width=\textwidth]{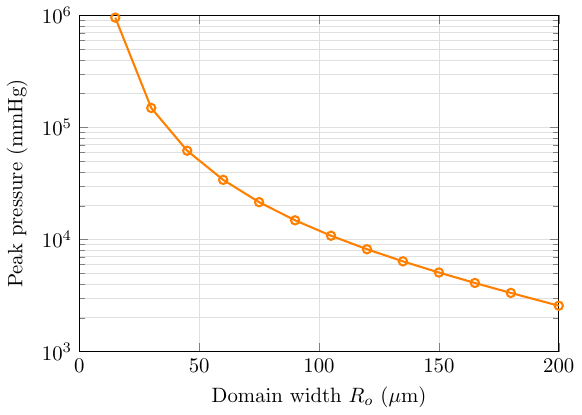}\caption{}\label{fig:ResHetPVSPPC1}}\end{subfigure}\hfill
    \begin{subfigure}{0.3\linewidth}{\includegraphics[width=\textwidth]{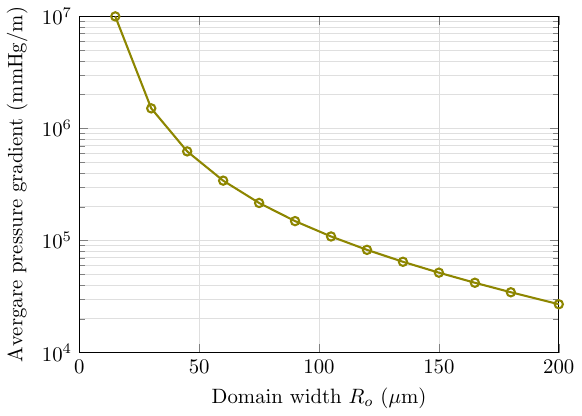}\caption{}\label{fig:ResHetPVSAPGC1}}\end{subfigure}
    
    \caption{Subproblem 3: numerical simulations of fluid flow problem in the \ac{PVS} and brain parenchyma with variable referential porosity.  
    \subref{fig:ResHetPVSVfzC1Ro15} Axial fluid velocity and \subref{fig:ResHetPVSVfltzC1Ro15} axial filtration velocity for $R_o = \np[\mu m]{15}$. 
    \subref{fig:ResPVSVfltzC1Ro15} Axial filtration velocity under uniform referential porosity ($\phi_{R_\f} = \np{0.2}$) with $R_o = \np[\mu m]{15}$.  
    \subref{fig:ResHetPVSVfzC1Ro200} Axial fluid velocity and \subref{fig:ResHetPVSVfltzC1Ro200} axial filtration velocity for $R_o = \np[\mu m]{200}$. 
    \subref{fig:ResPVSVfltzC1Ro200} Axial filtration velocity under uniform referential porosity ($\phi_{R_\f} = \np{0.2}$) with $R_o = \np[\mu m]{200}$.  
    \subref{fig:ResHetPVSVFRC1} Volume flow rate, \subref{fig:ResHetPVSPPC1} peak pressure, and \subref{fig:ResHetPVSAPGC1} average pressure gradient as a function of the outer radius $R_o$. The radial and axial axes are scaled differently to enhance visualization.}
    \label{fig:ResHetPVS}
\end{figure}

Figure~\ref{fig:ResHetPVS} presents the axial component of the fluid velocity and filtration velocity, as well as the volume flow rate, peak pressure, and average pressure gradient as functions of the outer radius.
Similarly to the previous case with uniform referential porosity, the flow rate, peak pressure, and average pressure gradient all exhibit a rapid decrease as the radius of the region increases, as illustrated in Figs.~\ref{fig:ResHetPVS}\subref{fig:ResHetPVSVFRC1}--\subref{fig:ResHetPVSAPGC1}.
Additionally, it is noted that the flow rates for $R_o=\np[\mu m]{15}$ and $R_o=\np[\mu m]{200}$ are $\np[\mu m^3/s]{2.89e5}$ and $\np[\mu m^3/s]{3.96e3}$, respectively, which are found to be comparable to the flow rates observed in the uniform referential porosity case.
The observed behavior can be attributed to the distributions of axial fluid velocity and axial filtration velocity, as presented in Figs.~\ref{fig:ResHetPVS}\subref{fig:ResHetPVSVfzC1Ro15}--\subref{fig:ResPVSVfltzC1Ro200}. 
Both axial fluid velocity and filtration velocity distributions are directly influenced by variable porosity, leading to distributions that differ from those observed in the constant porosity scenario.
Furthermore, the peak values of axial filtration velocities are comparable across cases, as shown in Figs.~\ref{fig:ResHetPVS}\subref{fig:ResHetPVSVfltzC1Ro15}--\subref{fig:ResPVSVfltzC1Ro15} and~\ref{fig:ResHetPVS}\subref{fig:ResHetPVSVfltzC1Ro200}--\subref{fig:ResPVSVfltzC1Ro200}. This similarity is primarily due to the relatively narrow width of the fluid-rich lumen (high porosity region) surrounding the penetrating artery in relation to the overall lumen width of the parenchyma, which ensures that the flow within this region does not substantially affect the total flow rate.

\subsection{Discussion}\label{sec: discussion}
Flow within the \ac{PVS} driven by peristalsis has been previously analyzed \cite{Wang2011Fluid-Mechanics-0} using simplified Darcy flow equations formulated for a moving periodic domain. An important assumption in \ac{WO} is that the porosity is not affected by deformation. This approach evaluated the contribution of pressure gradients and peristaltic mechanisms in facilitating \ac{CSF} movement along the \ac{PVS} surrounding a penetrating artery.
To incorporate the effects of structural deformation in the \ac{PVS}, we utilize a poroelastic model to study the fluid dynamics. Our findings indicate that employing baseline geometric parameters---specifically, a peristaltic wave amplitude of $\np[\mu m]{0.25}$, a referential porosity of $\np{0.2}$, and a lumen width of $\np[\mu m]{1}$---along with other key physical parameters, results in a critical scenario where the porosity effectively drops to zero. In practical terms, the deformation caused by the peristaltic wave, which amounts to $\sfrac{1}{4}$ of lumen width, is sufficient to expel nearly all fluid from some regions of the domain. Traditional computational models are typically unable to provide solutions in these kind of limiting situations. This finding provided the original impetus to develop the approach presented herein.

When brain parenchyma deformation is included in our model, our computations reveal that the resulting flow rates and pressure levels decrease, though they remain far from physiologically realistic ranges. Moreover, our findings indicate that peristaltic movements generate pressure gradients capable of driving \ac{CSF} flow both within the \ac{PVS} and into the surrounding parenchymal tissue.
The simplicity of our computational framework precludes an exhaustive exploration of transport dynamics. Nevertheless, this work provides critical insights into ongoing investigations of cerebral transport mechanisms. A key conclusion from our analysis is that tissue deformation constitutes a critical factor in governing fluid motion within the \ac{CNS}. The experiments using conventional Darcy flow formulations fail to capture fluid expulsion from the system, a phenomenon arising from bidirectional fluid-structure coupling between deformable tissues and interstitial fluid. These results underscore the inadequacy of simplified fluid models, such as Darcy, in resolving the intricate mechanics of cerebral fluid-structure interactions. Consequently, advanced poroelastic modeling approaches, which explicitly account for coupled solid-fluid dynamics, emerge as indispensable tools for rigorous investigation of this multiphysics phenomenon.

\section{Summary}
In this paper, we developed a poroelastic model based on mixture theory that can handle extreme conditions, i.e., the complete expulsion of the fluid component from certain regions of the domain. After verifying the proposed formulation using the method of manufactured solutions, we reformulated the poroelastic model in a moving frame to study the fluid mechanics within the \ac{PVS} surrounding a penetrating artery. Our findings revealed that the parameters suggested by \acl{WO} \cite{Wang2011Fluid-Mechanics-0} for the fluid flow in the \ac{PVS} lead the model to a zero porosity case which has not been addressed in previous investigations. Furthermore, considering the deformation of the brain tissue is crucial to obtain more physiologically reasonable results, which requires the use of a poroelastic model rather than the traditional Darcy model.

\appendix
\section{Mesh refinement study}

To determine the optimal mesh refinement for the computational domain of the \ac{PVS} fluid flow problem described in Section \ref{sec: PVS}, we analyzed four progressively refined computational meshes with characteristic element sizes of $\np[mm]{20}$, $\np[mm]{10}$, $\np[mm]{1}$, and $\np[mm]{0.1}$, corresponding to $\np{5}$, $\np{10}$, $\np{25}$, and $\np{50}$ elements in the radial direction, respectively. 
\begin{figure}[hbt]
    \centering
    \includegraphics[width=0.6\linewidth]{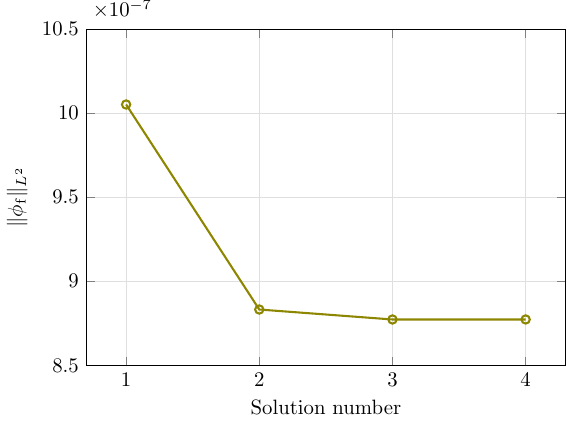}
    \caption{The $L^2$-norm of the porosity which is demonstrated in the deformed configuration of the computational domain. The norm converges to a constant value as the mesh refines.}
    \label{fig:porosity_norm}
\end{figure}
The $L^2$-norm of the porosity over the deformed configuration of the domain is presented in Fig.~\ref{fig:porosity_norm}. 
This norm serves as an indicator of the sufficiency of the refinement level for the numerical problem. The relative error in the $L^2$-norm of the porosity between the third and fourth refinement levels is less than $\np{2.5e-5}$. Consequently, all the results were computed using a mesh with a characteristic element size of $h_e=\np[mm]{0.1}$ and $\np{50}$ elements in the radial direction.

\subsection*{Acknowledgements}
The authors would like to express their gratitude to Prof.~Patrick J.~Drew of the Penn State Center for Neural Engineering for his valuable feedback and suggestions.

\subsection*{Funding}
The authors gratefully acknowledge partial support from the Pennsylvania Department of Health using Tobacco CURE Funds (the Department specifically disclaims responsibility for any analyses, interpretations, or conclusions).

\bibliographystyle{elsarticle-num}
\bibliography{LimitCaseLiterature}
\end{document}